\documentclass[fleqn,10pt]{wlscirep}
\usepackage{rotating}
\usepackage{subfigure}

\graphicspath{{figures/}}

\title{Estimating the epidemic risk using non-uniformly sampled contact data}

\author[1]{Julie Fournet}
\author[1,2,*]{Alain Barrat}

\affil[1]{Aix Marseille Univ, Universit\'e de Toulon, CNRS, CPT, UMR 733
2, 13288}
\affil[2]{Data Science Laboratory, ISI Foundation, Torino, Italy}
\affil[*]{alain.barrat@cpt.univ-mrs.fr}

\begin{abstract}
Many datasets describing contacts in a population suffer from
incompleteness due to population sampling and underreporting of
contacts.  Data-driven simulations of spreading processes using such
incomplete data lead to an underestimation of the epidemic risk, and
it is therefore important to devise methods to correct this bias. We
focus here on a non-uniform sampling of the contacts between
individuals, aimed at mimicking the results of diaries or surveys, and
consider as case studies two datasets collected in different contexts.
We show that using surrogate data built using a method developed in
the case of uniform population sampling yields an improvement with
respect to the use of the sampled data but is strongly limited by the
underestimation of the link density in the sampled network. We 
put forward a second method to build surrogate data that assumes
knowledge of the density of links within one of the groups forming the
population. We show that it gives very good results when the
population is strongly structured, and discuss its limitations in the
case of a population with a weaker group structure. These limitations highlight the
interest of measurements using wearable sensors able to yield accurate information on the structure and durations
of contacts.
\end{abstract}

\begin{document}

\flushbottom
\maketitle

\section*{Introduction}

An increasing number of studies on epidemic spreading processes use data-driven models. In particular, contact patterns between individuals
are considered to play an important role in determining the possible outcome of the transmission of infectious diseases in a population
\cite{Eames:2015,Voirin:2015,Obadia:2015}. Many datasets describing contacts between individuals in various contexts have
thus been gathered by different research groups, using techniques ranging from surveys or diaries to wearable sensors 
\cite{Pentland:2008,Mossong:2008,Mikolajczyk:2008,Cattuto:2010,Salathe:2010,Conlan:2011,Smieszek:2012,Hornbeck:2012,Read:2012,Stopczynski:2014,Mastrandrea:2015,Barrat:2015,Toth:2015,Guclu:2016}. The resulting data are typically in the form of contact networks in which nodes represent individuals
and edges represent the existence of at least one contact between the individuals linked.

Such network data can however be incomplete, for two main reasons.
On the one hand, not all individuals agree to participate to the data collection (either not answering the surveys or not willing to wear a sensor),
leading to node sampling. On the other
hand, contacts between participating individuals might not all figure in the gathered data, so that links are missing
from the data. In the case of diaries for instance, each individual
only remembers a fraction of his/her contacts, the longest contacts being better reported \cite{Smieszek:2014,Mastrandrea:2015,Smieszek:2016}.
If instead the available data comes from a survey about friendship relations, 
it will typically miss many short encounters between people who are not friends \cite{Mastrandrea:2015}.
Finally, even in the case of wearable sensors, some short contacts might not be 
detected, the actual detection of a contact might depend on its duration, depending on the sensitivity of the measuring
infrastructure, and the temporal
resolution may vary \cite{Stopczynski:2015}. It is thus of interest to understand how the resulting data incompleteness or limited resolution affects the properties of the measured 
contact network \cite{Lee:2006,Kossinets:2006,Genois:2015,Stopczynski:2015}, 
how it affects the outcome of data-driven models using incomplete data \cite{Ghani:1998,Genois:2015,Stopczynski:2015,Vestergaard:2016}, 
and most importantly if it is possible to 
infer the real network structure or statistical properties from incomplete information 
\cite{Bliss:2014,Zhang:2015,Squartini:2017} and/or to devise methods to 
correctly estimate the epidemic risk even from incomplete data \cite{Genois:2015,Mastrandrea:2016}
[Note that, for some types of wearable sensors, the opposite problem of false
positives, i.e., of reported contacts that are not relevant for propagation events, can also arise. Here we focus
on data incompleteness, but investigations of the impact of false positives would also be of clear interest]. 
To obtain such an estimation, a possibility is to try and construct surrogate datasets using
only the information contained in the incomplete data such that these surrogate data, despite not being strictly
equal to the original data, are ``similar enough'' with respect to the spreading process of interest. Here, ``similar enough'' means
that the outcomes of simulations of spread using the surrogate data should be close to the ones using the real, complete  data.

This issue has been addressed in the case of uniform population sampling by G\'enois et al. \cite{Genois:2015}. Uniform node sampling indeed
maintains not only the network density, but also the whole contact matrix of densities that describes the structure of links
in a population structured in distinct groups, such as classes in a school. We recall that the density of a network of $N$ nodes and $E$ edges
is defined as the ratio of the number of edges to the maximal possible number of edges that could exist between the nodes, i.e., $d = E / (N(N-1)/2)$. 
The contact matrix of densities gives for each pair of groups $X$ and  $Y$ the ratio between 
 the total number of links $E_{XY}$ between the $n_X$ individuals in $X$ and the $n_Y$ individuals in $Y$, and
the maximum number of such possible links ($n_Xn_Y$ if $X \ne Y$ or $n_X(n_X-1)/2$ if $X=Y$).
Moreover, 
the sampling also maintains the temporal statistics of contacts. It is thus
possible to measure this contact matrix and the temporal statistics in the incomplete data
and to construct surrogate data having the same statistics as the original one. Spreading processes simulated using
such surrogate data have been shown
to reproduce well the outcome of simulations using the whole dataset \cite{Genois:2015}. A similar method has been shown to work well also 
in a case study of contact diaries collected together with data from wearable sensors \cite{Mastrandrea:2016}: although not all contacts were 
reported in the diaries, building surrogate data using the contact matrix measured in the diaries and publicly available
statistics on contact durations made it possible to correctly estimate the outcome of simulations of spreading processes. 
In these two studies, the density of the sampled data was however either equal (for uniform population sampling) or close 
(for the diaries) to the one of the original data. In other cases, such as e.g. the friendship survey of Ref.~\cite{Mastrandrea:2015}, the 
density of the data is much smaller than the one of the contact data, and the method has indeed been shown to fail in this case \cite{Mastrandrea:2016}.
Sampled data with smaller density than the original one occur as soon as the links between sampled nodes are not all present.
This is expected to be the case in particular for data coming from diaries or surveys, and it is of interest to test the limits
of the reconstruction method, as a function of the sampling properties, and possibly to understand how to overcome this obstacle.

Here, we tackle this issue by considering incomplete contact data stemming from a non-uniform sampling procedure intended to mimic
a survey procedure in which (i) not all individuals participate and (ii) the contacts of each respondent are reported with a probability
depending on their duration \cite{Fournet:2016}. We consider empirical contact datasets, resample them using this non-uniform sampling 
procedure and design surrogate data as described above. We show that, at low sampling, the use of such surrogate
data in simulations of spreading processes is not enough to estimate the epidemic risk, even if it yields an improvement with respect
to the use of the raw sampled data. We thus consider the case in which additional information is available for one of the groups
forming the population: if one group is uniformly sampled (for instance if wearable sensors are available for this group), 
yielding a good estimate of its density, it is possible through a simple
rescaling procedure to estimate a new contact matrix for the data and to use it to construct another surrogate dataset that yields better results.
By using two datasets with different structures and varying the parameters of sampling and spreading processes, we explore 
the efficiency and limits of both procedures.

\section*{Data and methodology}

\begin{figure}[ht]
        \centering
        \includegraphics[width=0.45\textwidth]{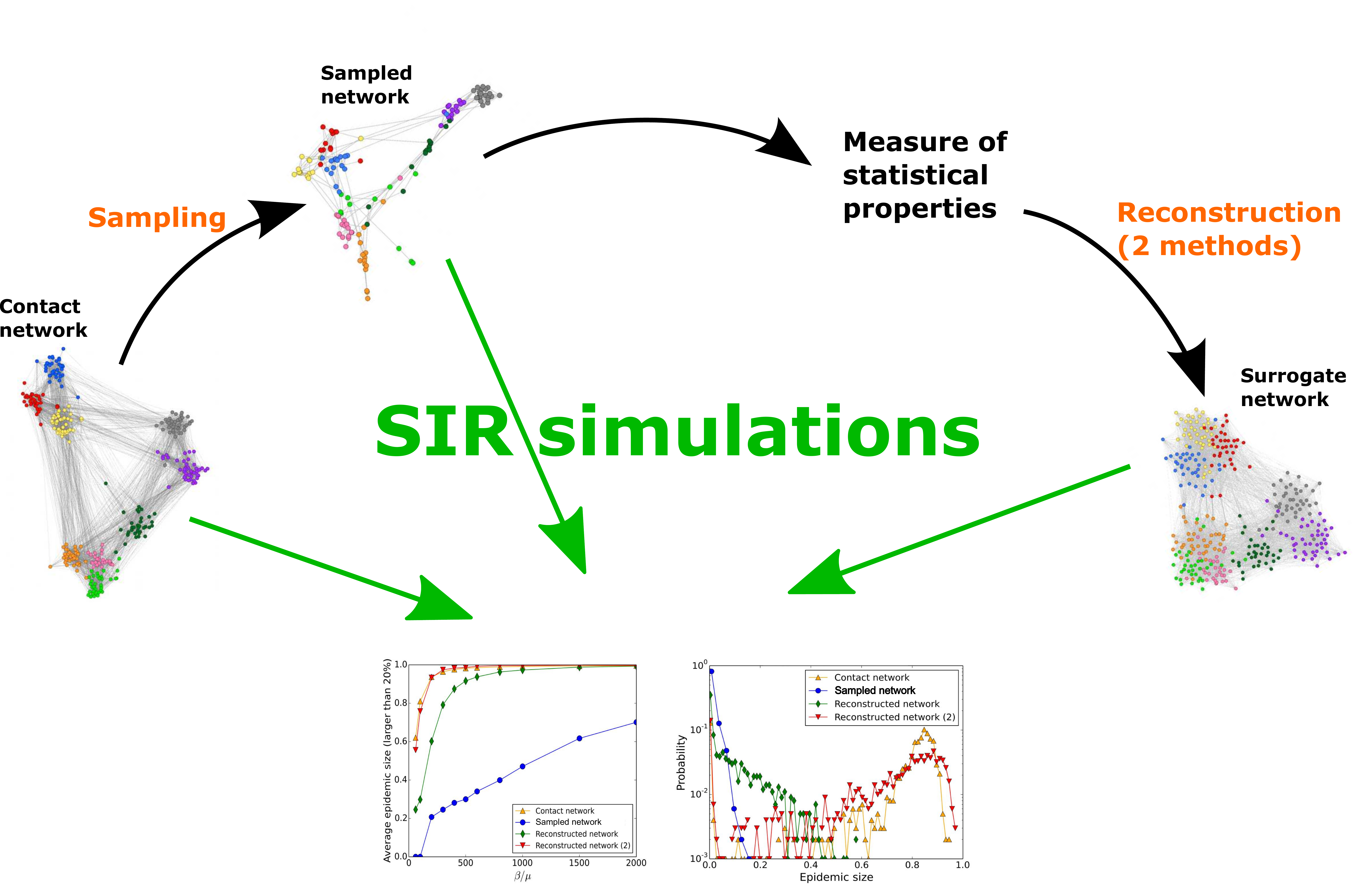}
\caption{\textbf{Sketch of the procedures considered in the article.} We 
consider a dataset describing a contact network. We perform a resampling
according to a certain (non-uniform) sampling method. We then measure a number of statistics of 
the resampled data and use these statistics to construct surrogate data. We simulate spreading processes
on the original, resampled and surrogate data and compare the outcomes, as given by the fraction of large
epidemics and by the whole distribution of epidemic sizes.
        \label{fig:sketch0}}
\end{figure}

We first describe the datasets and the different steps of our methodology, which are in the same spirit
as Refs~\cite{Genois:2015,Mastrandrea:2016}. Starting from a network of contacts between individuals
in a population, we perform a specific resampling that leads to an incomplete dataset. We describe two methods to create surrogate datasets
using the statistical information contained in the incomplete data. Spreading processes are then simulated on top of
the incomplete and of the surrogate datasets, and their outcomes compared with the ones of simulations using
the whole original contact network. The procedure is summarized in Fig. \ref{fig:sketch0}.

\subsection*{Data}

\begin{figure}[htb]
        \centering
        \includegraphics[width=0.45\textwidth]{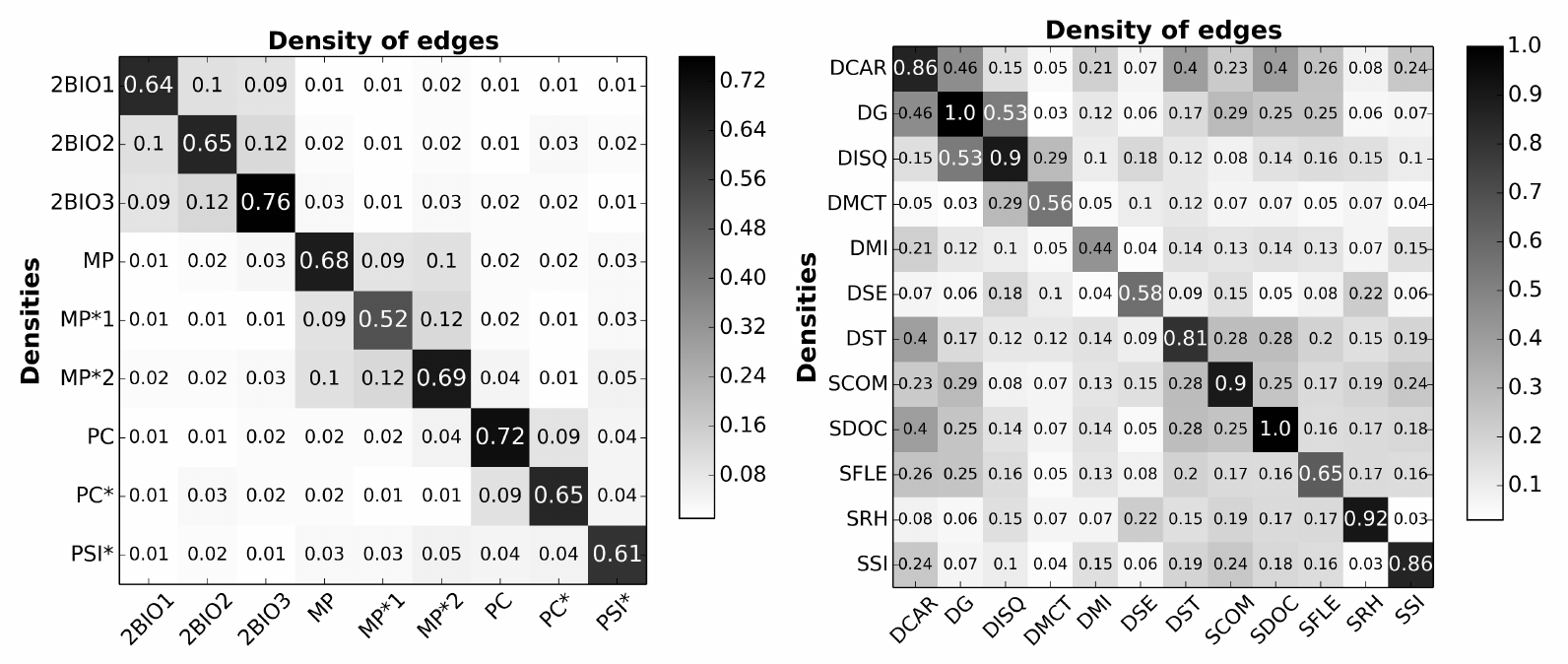}
        \caption{{\bf Density contact matrices.} Left: contact matrix giving the density of edges between classes during the study (Thiers13),
right: contact matrix giving the density of edges between departments during the study (InVS).
For each matrix, the entry at row $X$ and column $Y$ is given
by the total number of links between individuals in class or department $X$ and individuals in class or department $Y$,
  normalized by the maximum number of observable links ($n_Xn_Y$ or $n_X(n_X-1)/2$ if $X=Y$,
  with $n_X$ the cardinality of $X$). These matrices give only structural information as they do not take into account edge weights.
}
        \label{fig:dens_mat}
\end{figure}

We will use two datasets describing face-to-face contacts between individuals, 
collected and made publicly available by the
SocioPatterns collaboration (see the SocioPatterns website
http://www.sociopatterns.org/). 
The first dataset (Thiers13) has been collected in a French high school in December 2013. The
resulting contact network has $N=327$ nodes representing students (divided in $9$ classes corresponding to different
fields of study) and $E=5818$ weighted edges (see \cite{Mastrandrea:2015} for a detailed description and analysis of this dataset), for an average
link density of $2E/(N(N-1)) \approx 0.11$. 
The participation rate reached $86.5\%$ (there were overall $378$ students in the 9 classes).
The classes are of similar sizes (see Table \ref{tab1}), and
most edges ($69\%$, accounting for $93\%$ of the weights, i.e., of the total contact time between students) are found within classes (Fig. \ref{fig:dens_mat}).
The second dataset (InVS) has been collected in the office buildings of the Institut de Veille
Sanitaire (French Institute for Public Health Surveillance) in March
2015, and describes the contacts between $217$ individuals divided in $12$ departments. The contact network has $4274$ edges, i.e.,
a density of $\approx 0.18$. The departments 
have very different sizes and participation rate, with an average participation rate of $60\%$
(see Table \ref{tab2}) and, while most contact time occurs within departments ($76\%$), the corresponding fraction of edges
is only $42\%$ (Fig. \ref{fig:dens_mat}). 
In each dataset, an edge between two individuals corresponds to the fact that these individuals
have been in contact at least once during the data collection, and the edge weight gives the total contact time
between them. The contact matrices of edge densities are shown in Fig. \ref{fig:dens_mat} for both datasets.
In the Thiers13 case, we have moreover access to a network describing friendship relations between students,
obtained by a survey to which $135$ of the $327$ students answered. This friendship network has $413$ unweighted edges.

\begin{table}[htb]
\begin{tabular}{|l|c|c|c|c|}
	\hline
	Classes & Number of individuals & \% of the total population  & Participation rate & Number of participants to the friendship survey\\
	\hline
	2BIO1 & 36 & 11\% &  92\% & 10\\
	2BIO2 & 34 & 10.4\% &  81\% & 20\\
	2BIO3 & 40 & 12.2\%&  98\% & 28\\
	PC & 44 & 13.5\% & 98\% & 21\\ 
	PC* & 39 & 12\%  & 95\% & 10\\
	PSI* & 34 & 10.4\%  & 77\% & 15\\
	MP & 33 & 10\% & 79\% & 21\\
	MP*1 & 29 & 8.9\% & 69\% & 3\\
	MP*2 & 38 & 11.6\%  & 90\% & 7\\
	Total & 327 & 100\% & 86\% & 135\\
	\hline
\end{tabular}
\caption{Number of individuals in each class participating to the data collection in the highschool (Thiers13 dataset), percentage
with respect to the population under study and participation rates. 
	\label{tab1}}
\end{table}

\begin{table}[htb]
\begin{tabular}{|l|c|c|c|}
	\hline
	Departments & Number of individuals & $\%$ of the total population  & Participation rate\\
	\hline
	DCAR & 13 & 6\%   & 65\% \\
	DG & 2 &  0.9\% & 22\%  \\
	DISQ & 18 &  8.3\%  & 86\% \\
	DMCT & 31 &  14.3\%  & 63\% \\
	DMI & 57 &  26.3\%  & 79\% \\
	DSE & 32 & 14.7\% & 56\% \\
	DST & 23 &  10.6\% & 52\% \\
	SCOM & 7 & 3.2\% &  70\% \\
	SDOC & 4 &  1.8\% &  57\% \\
	SFLE & 14 &  6.5\% & 37\% \\
	SRH & 9 &  4.2\% & 64\% \\
	SSI & 7 &  3.2\% &  32\% \\
	Total & 217 & 100\% & 60\% \\
	\hline
\end{tabular}
\caption{Number of individuals in each department 
participating to the data collection in the office buildings of InVS, percentage with respect to the population under study and participation rate.
	\label{tab2}}
\end{table}

\subsection*{Sampling method}

We consider the sampling method put forward in \cite{Fournet:2016}, called ``EGOref'', based on 
(i) a uniform sampling of nodes and (ii) a non-uniform sampling of edges, edges with larger weights 
being preferentially sampled. This procedure is designed to mimic a sampling of links obtained for instance through 
a survey on friendship relations in a population: it is 
inspired by the result of Ref. \cite{Mastrandrea:2015} that the longest
contacts measured in the Thiers13 dataset corresponded to reported friendships, while many short contacts did not.
In particular, we have shown in Ref. \cite{Fournet:2016} that the outcome of simulations of spreading processes on 
the friendship network of the Thiers13 dataset can be reproduced if the EGOref sampling method 
is applied to the contact network of the same dataset, with correctly adjusted sampling parameters.

\begin{figure}[ht]
        \centering
        \includegraphics[width=0.9\textwidth]{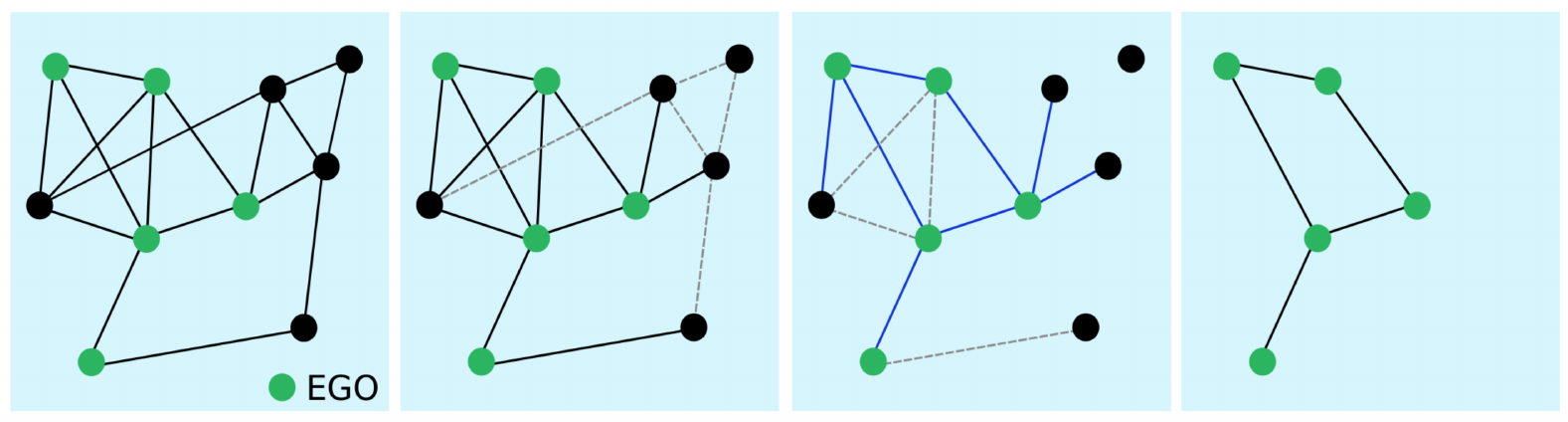}
\caption{\textbf{Sketch of the EGOref sampling process.} We first
  select a certain number of nodes as egos. Each ego ``chooses'' to
  report some of its links, with probability depending on their
  weights. A link can be selected twice if joins two \textit{egos}
  (blue edges). We then finally keep only the egos and, among
  the chosen edges, only the ones joining \textit{egos}.
        \label{fig:sketch}}
\end{figure}

More precisely, the EGOref sampling depends indeed on two
parameters. Starting from a weighted contact network with $N_0$ nodes,
we select $N$ of these nodes (called ``\textit{egos}'') uniformly at
random (a non-uniform selection could also be considered). For each \textit{ego} $i$, each edge $i-j$ is selected with a probability equal to
$p*\frac{W_{ij}}{S_{i}}$, with $W_{ij}$ the weight of the edge between
$i$ and $j$, $S_i = \sum_\ell W_{i\ell}$ the strength of the
\textit{ego} node $i$ and $p$ the sampling parameter. We then keep only the \textit{egos} and the
selected edges linking them and we remove the other edges (between egos and non-egos and between non-egos) and nodes (non-egos). 
With this method, we end up with a tunable number of nodes $N$ and a number of edges that
depends on the parameter $p$. Figure \ref{fig:sketch}
summarizes this process.

\begin{figure}[ht]
        \includegraphics[width=0.5\textwidth]{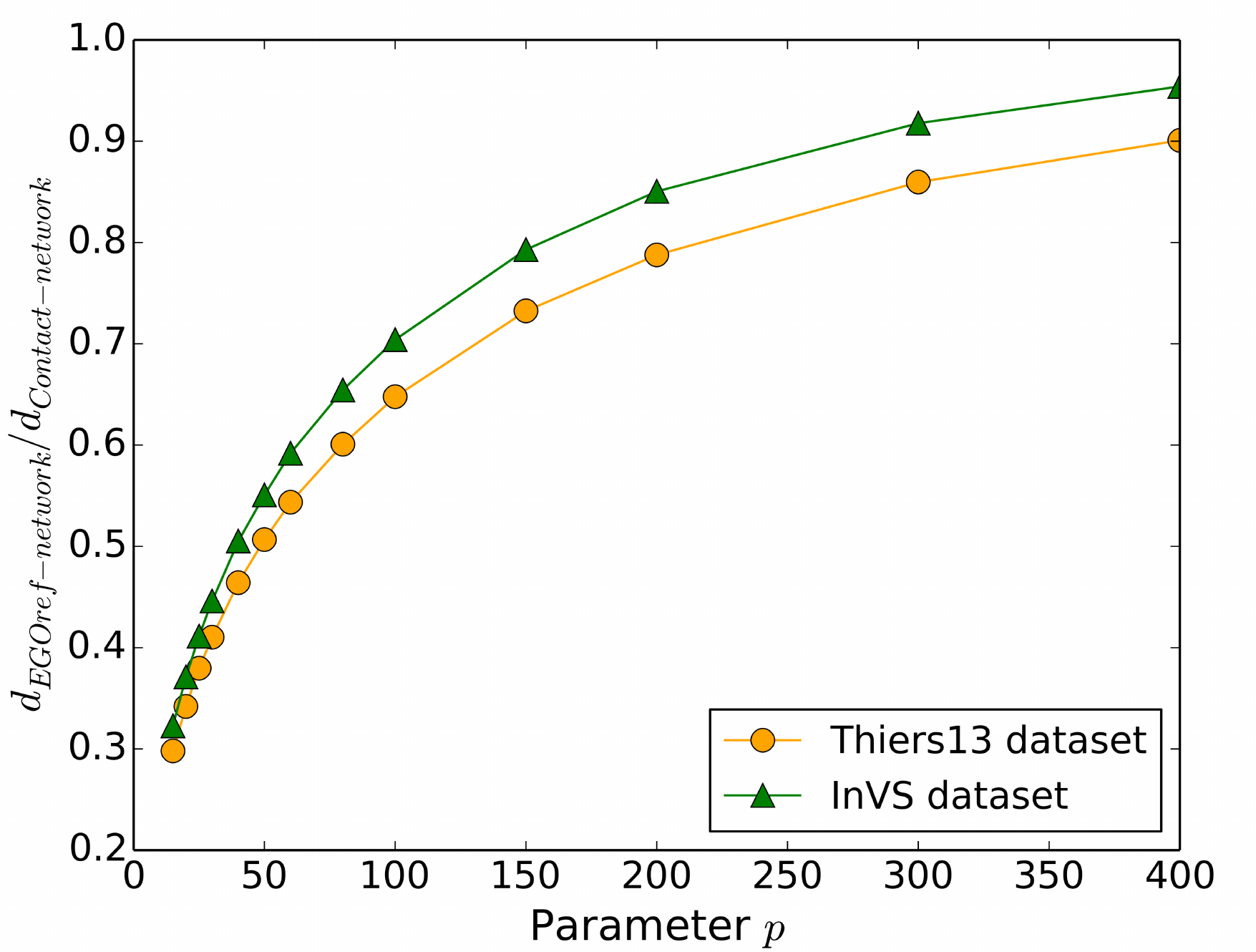}
        \caption{{\bf Impact of sampling on network density.} Ratio between the density of the EGOref sampled
          network and the density of the whole contact network as a
          function of the parameter $p$ for the two datasets. Here the
          number $N$ of sampled individuals is $70\%$ of the total
          population: changing this number does not change the ratio.}
        \label{fig:density}
\end{figure}

The parameter $p$ clearly has an effect on the density of the sampled network. This is in contrast with the
case of a uniform population sampling in which all the edges between sampled nodes are kept, and which thus conserves the network
density \cite{Genois:2015}. 
Figure \ref{fig:density} displays the ratio between the density of the EGOref sampled network and the original contact network 
for the two datasets used here: the density of the sampled network increases with the parameter $p$. On the other hand,
the density of the sampled network does not depend on the number of sampled nodes (Figure S1 in the Supplementary Information).
 
Despite the decrease in network density caused by the EGOref sampling, it is worth noting that the 
contact matrices of edge densities measured in the original contact network and 
in the EGOref sampled network remain very similar for both datasets (see Figures S2-S4 in the Supplementary Information).
These matrices give, for each pair of groups in the population (here, classes or departments), the number of links between these groups
normalized by the maximum possible number of such links (obtained if each member of one group is linked to all members of the other group). 
A large similarity between contact matrices indicates that the overall structure of the network is preserved by the sampling, even if
the specific values of densities are changed.

\subsection*{Surrogate data construction methods}

As in Refs. \cite{Genois:2015} and \cite{Mastrandrea:2016}, our initial goal is to use only the information contained in the sampled
data to construct surrogate contact networks that are statistically similar to the full data. 
As already made clear in the introduction and in refs \cite{Genois:2015,Mastrandrea:2016}, we emphasize again 
that the point is not to infer the missing links but to
build a ``plausible'' version of these links, such that the simulations of epidemic spread on the
resulting network, as described below, yield an accurate estimation of the epidemic risk. 
We consider two distinct methods to construct such surrogate data.

The first method is the equivalent for static networks of the one presented in \cite{Genois:2015,Mastrandrea:2016}. First, the contact
matrix of edge densities is measured in the incomplete data. Assuming that the number of missing nodes in each group (class or department)
is known, we add the missing nodes in each and we add links randomly in each group and between groups in such a way to keep the contact matrix
fixed to its measured value. 
More precisely, we first measure the density $d$ in the incomplete data as $2E/(N(N-1))$ where $N$ and $E$ are the numbers of nodes and edges
in these incomplete data. Knowing the number $n$ of missing nodes, we can deduce the number $e$ of additional links needed to keep the density constant 
when we add the $n$ missing nodes, through $d = 2(E+e)/((N+n)(N+n-1))$. We moreover
transform the contact matrix of edge densities $\rho_{XY}$ into a row-normalized contact matrix $C$, in which the element 
$C_{XY} = \rho_{XY} / \sum_Z \rho_{XZ}$ gives the probability for a node of group $X$ to have a link to a node of group $Y$. Then, 
for each missing edge, we proceed as follows: (i) we extract at random a node $i$ (among the total population of
$N+n$ nodes); (ii) knowing the group $X$ that $i$ belongs to, we extract at random a target group $Y$
with probability given by $C_{XY}$; (iii) we draw at random a node $j$ in group $Y$ such that $i$ and $j$ are not yet linked, and we add a link between $i$ and $j$.
The resulting network has the same number of nodes as the original network and the same
contact matrix of densities and overall density as the sampled network. Weights are finally assigned to the edges: weights
are taken at random from the empirical distribution of weights of the contact network, which is known to be a robust feature of human contact patterns
and does not depend on the context \cite{Barrat:2015}.
This method has been shown to yield good results when the incomplete data results from a uniform population
sampling of the original contact network \cite{Genois:2015}, which preserves the overall network density.
In Ref. \cite{Mastrandrea:2016}, it has also been shown to be able to create relevant surrogate data (i.e., yielding
the same outcome as the original data when used in simulations of spreading processes) from contact diaries data. Also in this latter case,
the density of the network deduced from diaries was similar to the one of the original contact data.
In Ref. \cite{Mastrandrea:2016} moreover, it has been shown that one can use a pool of publicly available contact duration statistics
to assign weights to the edges of the surrogate data.

However, as the EGOref sampling method yields sampled network with densities smaller than the original data (Fig. \ref{fig:density})
and as it is well known that density plays an important role in determining the outcome of a spreading process, we propose a second construction method, in which 
we assume in addition that the data includes the density of edges within one of the groups (chosen at random) in the original
 non-sampled network. We therefore first measure the contact matrix
 of edge densities $\rho_{XY}$ in the sampled data, as in the first
 method. We then compute the ratio $f$ between the real density
 $\tilde{\rho}_{AA}$ in the original data for the group $A$ that is
 assumed to be known and the measured density $\rho_{AA}$. 
 We then rescale the average density of the graph by the factor $f$, $d' = f \times d$ 
 and compute the new number $e'$ of edges to be added through $d' = 2(E+e')/((N+n)(N+n-1))$. 
 The procedure is then the same
 as the previous one: we add the missing nodes and $e'$ links in order to preserve the
 normalized contact matrix $C_{XY}$, and weights are taken from the empirical
 distribution of aggregated contact durations and assigned at random
 to edges.

The rationale behind the second method is that the infrastructure of
wearable sensors able to measure this density could have been
available for only one group in the population (or even a random
fraction of one group), for instance, while only partial information
from a survey is available for the other groups. We then assume that the
EGOref sampling method affects in a similar way all parts of the graph, so that a global rescaling
of the density, which also rescales all the elements of the contact matrix of edge densities by the same factor, should yield values closer to the original
data.

In the following, we will apply each method to contact networks sampled using the EGOref method for various values of the parameters
$p$ and $N$. As Ref. \cite{Fournet:2016} has shown the similarity between the friendship network of the Thiers13 dataset and
the outcome of the EGOref sampling applied to the Thiers13 contact network, for a specific parameter value, we will also apply
these methods to construct surrogate contact data using the friendship network instead of a sampled version of the original contact network.

\subsection*{Simulation of spreading processes}

Our goal is to understand if it is possible to use incomplete datasets to estimate the epidemic risk in a population
by constructing surrogate data and using them in the simulations of epidemic spread. As a paradigm of epidemic process, 
we consider the Susceptible-Infectious-Recovered (SIR) model: in this model, nodes are initially all susceptible (S), except 
one in the Infectious state, chosen at random and seed of the process. Each 
Susceptible (S) node $i$ can become infectious when in contact with an Infectious one $j$. This occurs at a rate 
$\beta W_{ij} /T$ where $W_{ij}$ is the weight of the link $i-j$ and $T$ the total measurement time \cite{Stehle:2011,Fournet:2016}
(i.e., the probability for $i$ to become infectious during a time step $dt$ is $\beta W_{ij} dt /T$). 
Infectious nodes become Recovered (R) at rate $\mu$ and cannot be
infected anymore. The process ends when there are no Infectious nodes any more. 

We perform numerical simulations of this model for each dataset on the original contact network, on the 
sampled networks at various values of the parameters $p$ and $N$, and on the surrogate datasets built using the two methods described above
(note that we will equivalently write ``surrogate data'' or ``reconstructed networks'' to describe the surrogate datasets).
For the Thiers13 case, we also perform simulations on the friendship networks and the corresponding surrogate data.
We also vary the ratio $\beta/\mu$ that modulates, in each given network, the impact of the modelled disease.

To quantify the epidemic risk, we measure in each simulation the epidemic size as given by the 
final fraction of recovered nodes. We compare the distributions of epidemic sizes, 
the fraction of epidemics with size larger than $20\%$ and the average size of these epidemics
(the cut-off of $20\%$ is chosen arbitrarily to distinguish between small and large epidemics; changing the value of
this threshold does not alter our results). 

We finally note that we consider here static versions of the contact networks, while the original SocioPatterns data 
provides temporally resolved contacts. The EGOref sampling process indeed mimics a procedure yielding 
a static sample of the actual contact network. In the context of models of infectious diseases with 
realistic timescales of several days, this can represent enough information to obtain an estimate of the epidemic risk, as discussed in Ref. \cite{Stehle:2011}.
For faster spread, one could add to the surrogate construction method an additional step of building realistic contact timelines
as in Ref \cite{Genois:2015}.

\section*{Results}

We discuss separately the results obtained with the two datasets. The Thiers13 one is indeed much more strongly ``structured'' than the InVS one, 
in the sense that the fraction of interactions occurring within each class is very high. Moreover, all classes are of similar sizes and have similar
link densities. Classes are also arranged in groups of $2$ or $3$ classes corresponding to the major topic of study of their students. In the InVS case, 
departments are of different sizes, their link densities vary more and the pattern of interactions between departments is less structured.

\begin{table}[htb]
	\begin{tabular}{|l|c|c|c|c|c|}
		\hline
		& N & E & d & Avg clustering & Avg shortest path \\
		\hline
		Contact network & 327 & 5818 & 0.11 & 0.503 & 2.15\\
		EGOref network & 135 & 405 & 0.04 & 0.351 & 3.95\\
		Reconstructed network 1 & 327 & 2386 & 0.04 & 0.263 & 3.12 \\
		Reconstructed network 2 & 327 & 4705 & 0.09 & 0.518 & 2.69 \\
		\hline
	\end{tabular}
	\caption{Thiers13 dataset: Features of the original contact network, of the sampled one and of the surrogate data, for an EGOref sampling with
$p=30$ and $N=135$. 
		\label{tab5}}
\end{table}

\begin{table}[htb]
\begin{tabular}{|l|c|c|c|c|c|}
	\hline
	 & N & E & d & Avg clustering & Avg shortest path \\
	\hline
	Contact network & 327 & 5818 & 0.11 & 0.503 & 2.15\\
	Friendship network & 135 & 413 & 0.05 & 0.532 & 4.06\\
	Reconstructed network & 327 & 5376 & 0.10 & 0.464 & 2.46 \\
	\hline
\end{tabular}
\caption{Thiers13 dataset: Basic features of the empirical networks and of the reconstructed network (obtained 
by applying the second method of construction of surrogate data to the friendship network).
	\label{tab3}}
\end{table}

\subsection*{First case: highly structured network (Thiers13 dataset)}

Table \ref{tab5} gives some basic features of the original, the EGOref sampled and the surrogate networks built with the
two methods described above, for the EGOref parameter values yielding number of nodes and edges similar to the friendship network.
This corresponds to $N=135$ and $p=30$. The sampled network has a much lower density than the original contact data, smaller
clustering and larger average shortest path. The surrogate network built using the first method has by construction the same density,
and has an even smaller clustering, while the second method yields values much closer to the original ones. Moreover, the similarities
between the contact matrices of the original, sampled and surrogate data all exceed $99\%$ (see also Figure S3 of the Supplementary Information).
 However, the fraction of intra-class links is larger in the reconstructed network than in the original one 
($83\%$ vs $69\%$, see Supplementary Information), 
while the fraction of contact durations these intra-class links carry is slightly
smaller ($83\%$ vs $93\%$; note that since weights are put at random,
the fraction of links and the fraction of weights they carry are the same in the reconstructed data).

\begin{figure}[htb]
        \centering
        \includegraphics[width=0.9\textwidth]{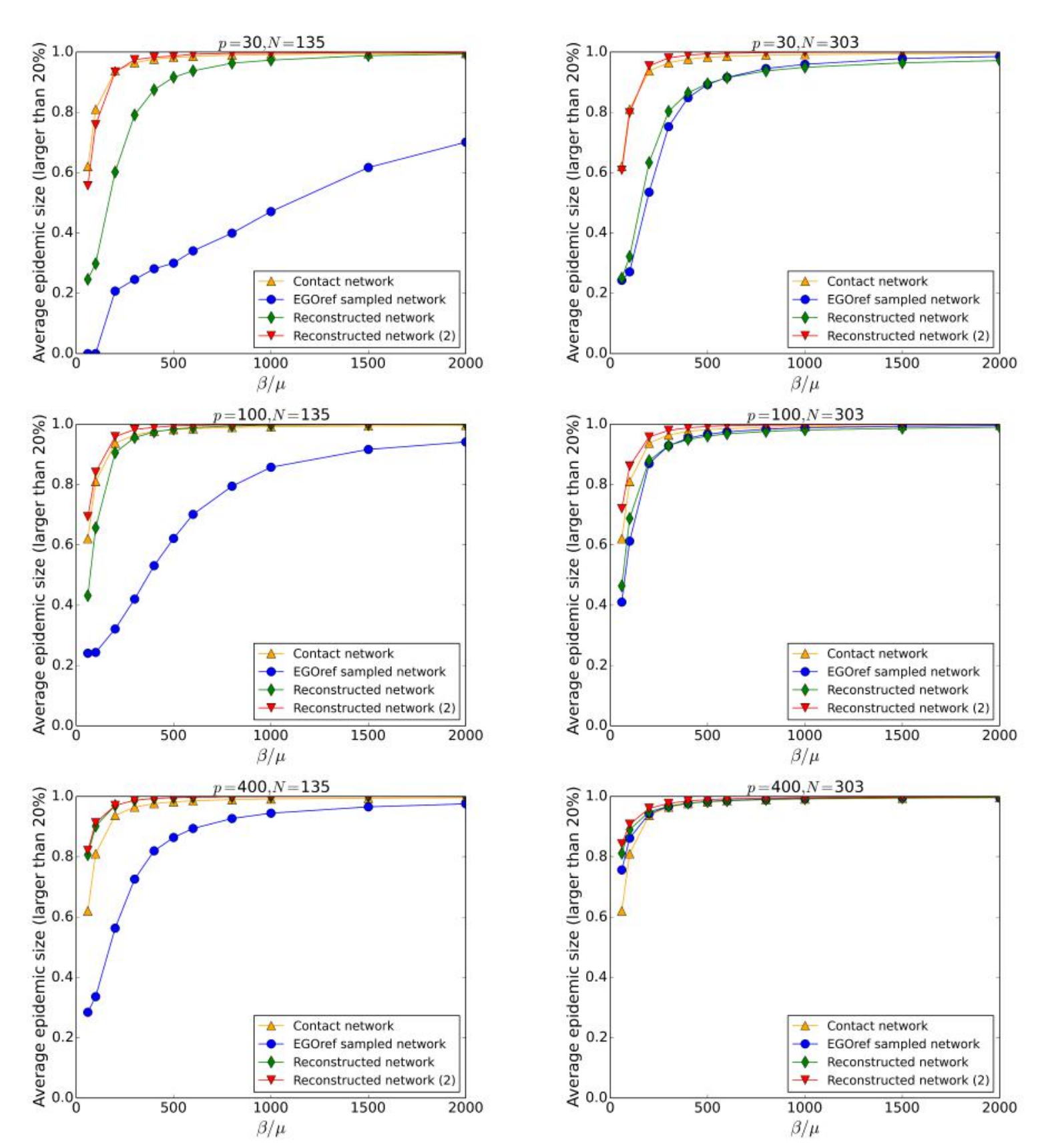}
        \caption{\textbf{Thiers13 dataset: Outcomes of SIR spreading simulations.} Average size of epidemics with size
above 20\% as a function of the spreading parameter $\beta/\mu$ for different values of $p$ and $N$.
The simulations are performed on contact network, EGOref sampled network and the reconstructed networks
using the two methods of reconstruction described in the text.}
        \label{fig:epi1_thiers}
\end{figure}

\begin{figure}[htb]
        \centering
        \includegraphics[width=0.8\textwidth]{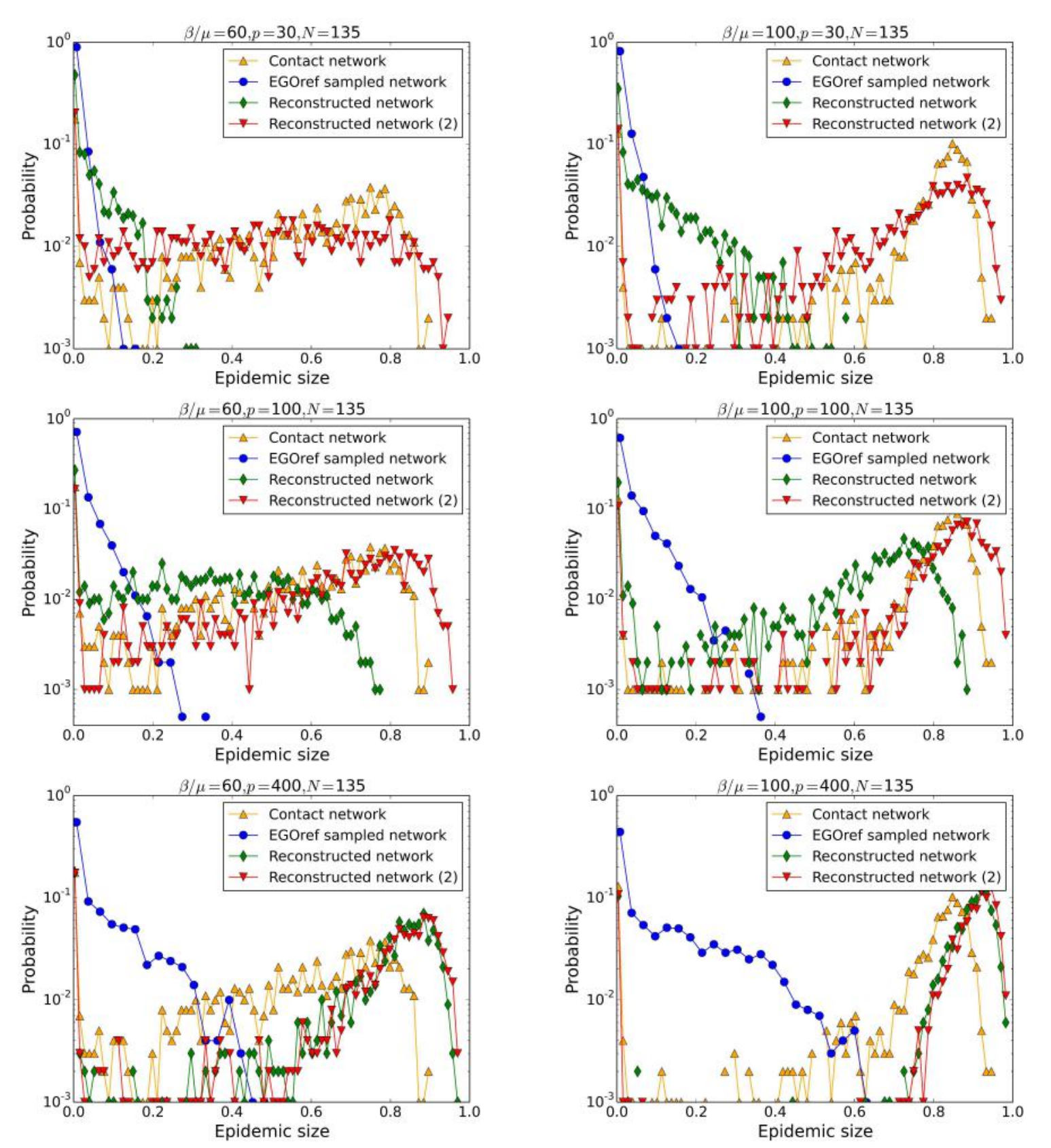}
        \caption{\textbf{Thiers13 dataset: Distributions of epidemic sizes for SIR spreading simulations.}
The simulations are performed on the contact network on the EGOref sampled network and on
the surrogate data (reconstructed networks) built using the two methods of reconstruction described in the text.
The parameter of spreading $\beta/\mu$, the parameter $p$ and the number of sampled individuals $N$ are given above each plot.
}
        \label{fig:epi2_thiers}
\end{figure}

Figures \ref{fig:epi1_thiers} and  \ref{fig:epi2_thiers}
compare the outcome of SIR simulations performed on the original contact network, on the EGOref sampled networks 
and on the surrogate data built using the two reconstruction methods, for various values of the sampling parameters.
Figure \ref{fig:epi1_thiers} first displays the average size of large epidemics 
(i.e., the ones reaching at least $20\%$ of the population)
as a function of the spreading parameter $\beta/\mu$. As expected and already explored \cite{Fournet:2016}, 
simulations performed on the EGOref sampled network yield a strong underestimation of the epidemic risk with 
respect to results obtained with the use of the contact network, except at large $p$ and $N$ (in this case, the sampled
network is almost equal to the original one, and the random assignment of weights to the links leads in fact to a 
slight overestimation of the epidemic risk \cite{Fournet:2016}; this occurs only at unrealistically large values of $p$).

The use of surrogate data obtained with the first method improves the estimation of the epidemic risk with respect to the use
of the sampled data but still leads to a clear underestimation for small and intermediate values of $p$. This is not unexpected
given the reconstruction method maintains the density of the sampled network. The second method, which leads to
surrogate data with densities closer to the original one, allows to obtain a much better estimation of the epidemic risk.

Figure \ref{fig:epi2_thiers} focuses on the case of $N=135$ and
displays the whole distributions of epidemic sizes obtained from SIR simulations
for different values of the spreading parameter $\beta/\mu$ and of the parameter of sampling
$p$. In all cases, the distributions obtained with the sampled network remain narrow and do not develop a peak at large 
values of epidemic sizes. The distributions obtained with surrogate data are both broader but can differ strongly from each other 
depending on the value of $p$. At very large $p$, the sampling procedure almost does not affect the network density,
so that both methods yield very close outcomes; as discussed above,  the random assignment of weights leads then to a peak
at large values of the epidemic size that is shifted to values larger than with the original network. 
For more realistic small and intermediate values of $p$, the first method leads to distributions that are much narrower
than the outcome of simulations on the original network, while the second method yields a much better agreement, albeit
with a systematic small overestimation of the largest epidemic sizes.

We now turn to the case of the friendship network. 
Table \ref{tab3} compares the main features of the contact network, the friendship network and the surrogate data
obtained by the second method of reconstruction applied to the friendship network. The reconstruction procedure
allows to recover a density similar to the one of the contact data. Moreover, the contact matrix of these three networks
are very similar (similarity values of more than $98\%$).
We however note that the fraction of within-classes links is larger 
in the friendship network ($75\%$) than in the contact network
($69\%$), and that this characteristics holds also for the surrogate data.

\begin{figure}[htb]
        \centering
        \includegraphics[width=0.8\textwidth]{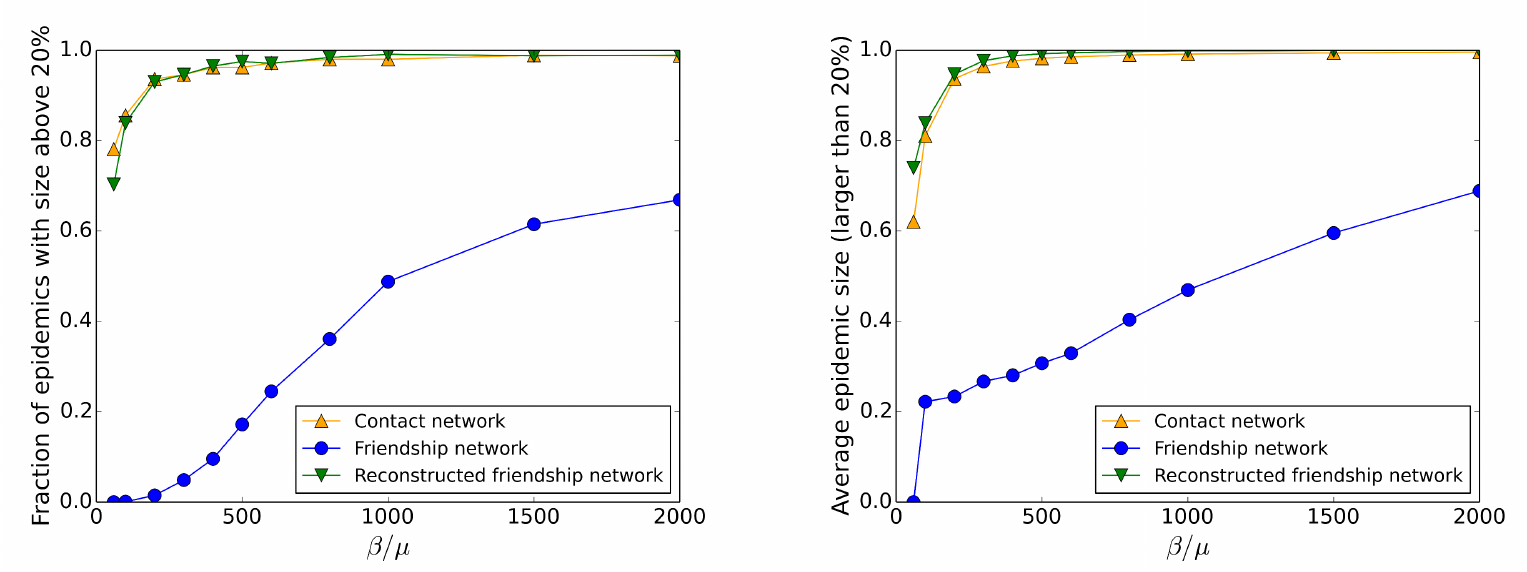}
        \caption{\textbf{Thiers13 dataset: Outcome of SIR spreading
            simulations.} Fraction of epidemics with size above 20\%
          as a function of the spreading parameter $\beta/\mu$ (left)
          and average size of epidemic with size above 20\% as a
          function of the spreading parameter $\beta/\mu$ (right). The
          simulations are performed on  the contact
          network, on the friendship network and on surrogate data obtained by applying the second method
of construction to the friendship network.}
        \label{fig:epi4_thiers}
\end{figure}

\begin{figure}[htb]
        \centering
        \includegraphics[width=0.8\textwidth]{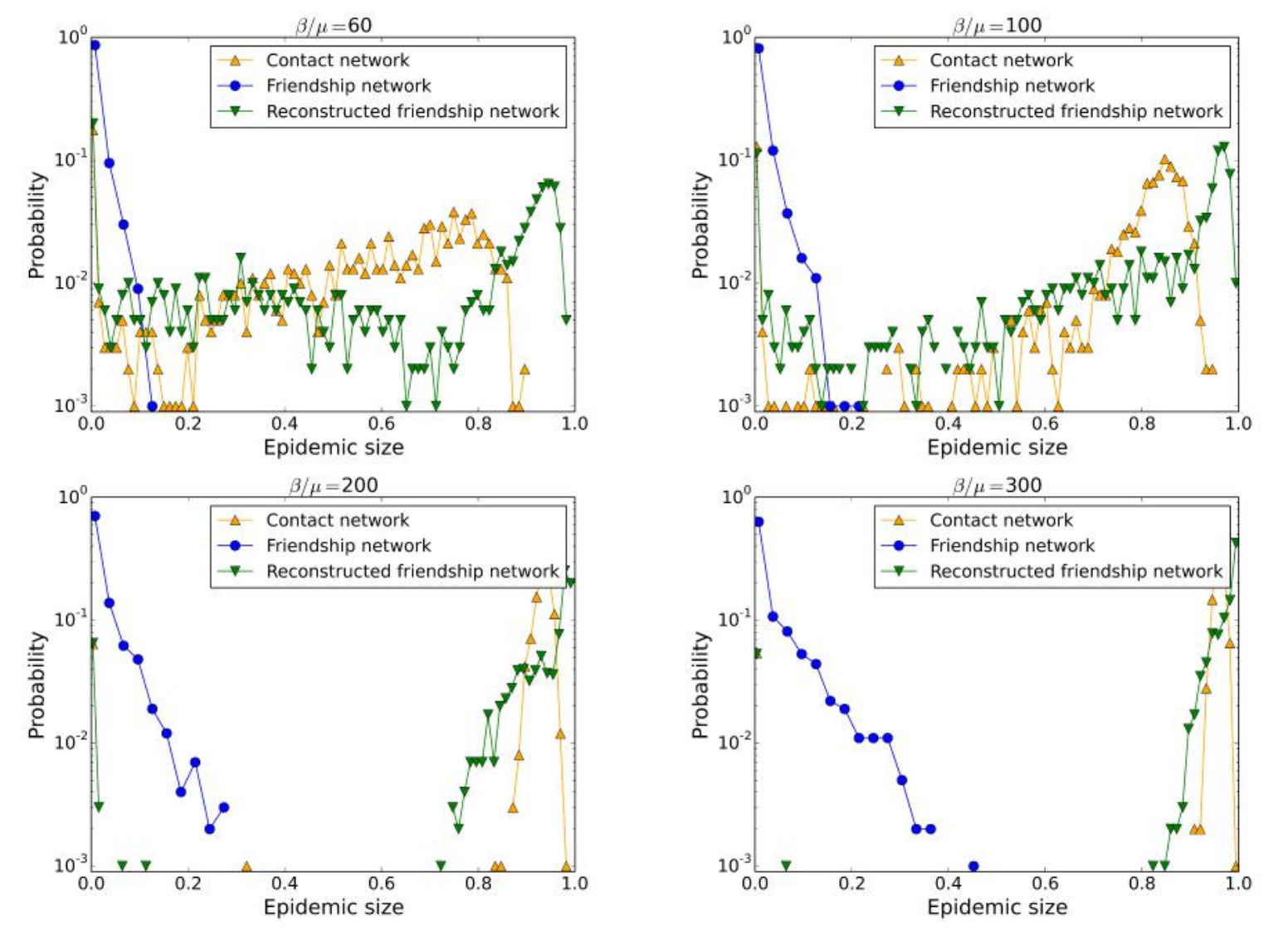}
        \caption{\textbf{Thiers13 dataset: Distributions of epidemic
            sizes of SIR spreading simulations.} The simulations are performed on the contact
          network, on the friendship network and on surrogate data obtained by applying the second method
of construction to the friendship network. The value of $\beta/\mu$ used is given
          above each plot.}
        \label{fig:epi3_thiers}
\end{figure}

Figures \ref{fig:epi4_thiers} and \ref{fig:epi3_thiers} show the
outcomes of epidemic spreading simulations performed on the empirical
networks and on the reconstructed network. While the simulations using
the friendship network leads to a strong under-estimation of the
epidemic risk, Figure \ref{fig:epi4_thiers} shows that the use of the
surrogate data yields a very good estimation of the fraction of
epidemics with size above $20\%$ and of the average epidemic size,
across a large range of values of $\beta/\mu$.  Figure
\ref{fig:epi3_thiers} displays the whole distributions of epidemic
sizes for different values of the spreading parameter $\beta/\mu$.  It
confirms that the surrogate data yields distributions with more
similar shapes to the contact network case than the friendship
network. However, the maximal sizes of epidemics are systematically overestimated, which might be ascribed to the larger
fraction of weights on inter-class links in the surrogate data with respect to
the contact network.

\subsection*{Second case: a less structured network (InVS dataset)}

We now investigate the results obtained with the InVS dataset. As discussed above, the structuration in departments 
leads to a less structured
contact matrix than in the highschool case. Table \ref{tab4} and Figure S4 of the Supplementary Information  compare some characteristics
of the sampled and surrogate data (with the second method of reconstruction) to the original network for $p=30$ and $N=93$
(i.e., a similar fraction of the population as used in the example of the Thiers13 dataset). Here, even the second method of
reconstruction leads to a network density smaller than the original one, even if much closer than for the sampled data.
The contact matrices of the sampled and of the surrogate data are very similar to the one of the original data 
(Figure S4 of the Supplementary Information), but, as in the previous case, the fraction of intra-department edges is larger
in the surrogate data than in the original one ($58\%$ versus $42\%$) while the fraction of the weights these links carry is 
much smaller ($58\%$ versus $76\%$ in the original contact data).

\begin{table}[htb]
	\begin{tabular}{|l|c|c|c|c|c|}
		\hline
		& N & E & d & Avg clustering & Avg shortest path \\
		\hline
		Contact network & 217 & 4274 & 0.18 & 0.38 & 1.88\\
		EGOref network & 93 & 348 & 0.08 & 0.27 & 2.64\\
		Reconstructed network & 215 & 3048 & 0.13 & 0.32 & 2.07 \\
		\hline
	\end{tabular}
	\caption{InVS dataset: Basic features of the contact network, of the EGOref sampled network (here with 
$p=30$ and $N$=40\% of the total number of nodes) and of the surrogate data obtained using 
the second method of reconstruction. 
		\label{tab4}}
\end{table}

\begin{figure}[htb]
        \centering
        \includegraphics[width=0.8\textwidth]{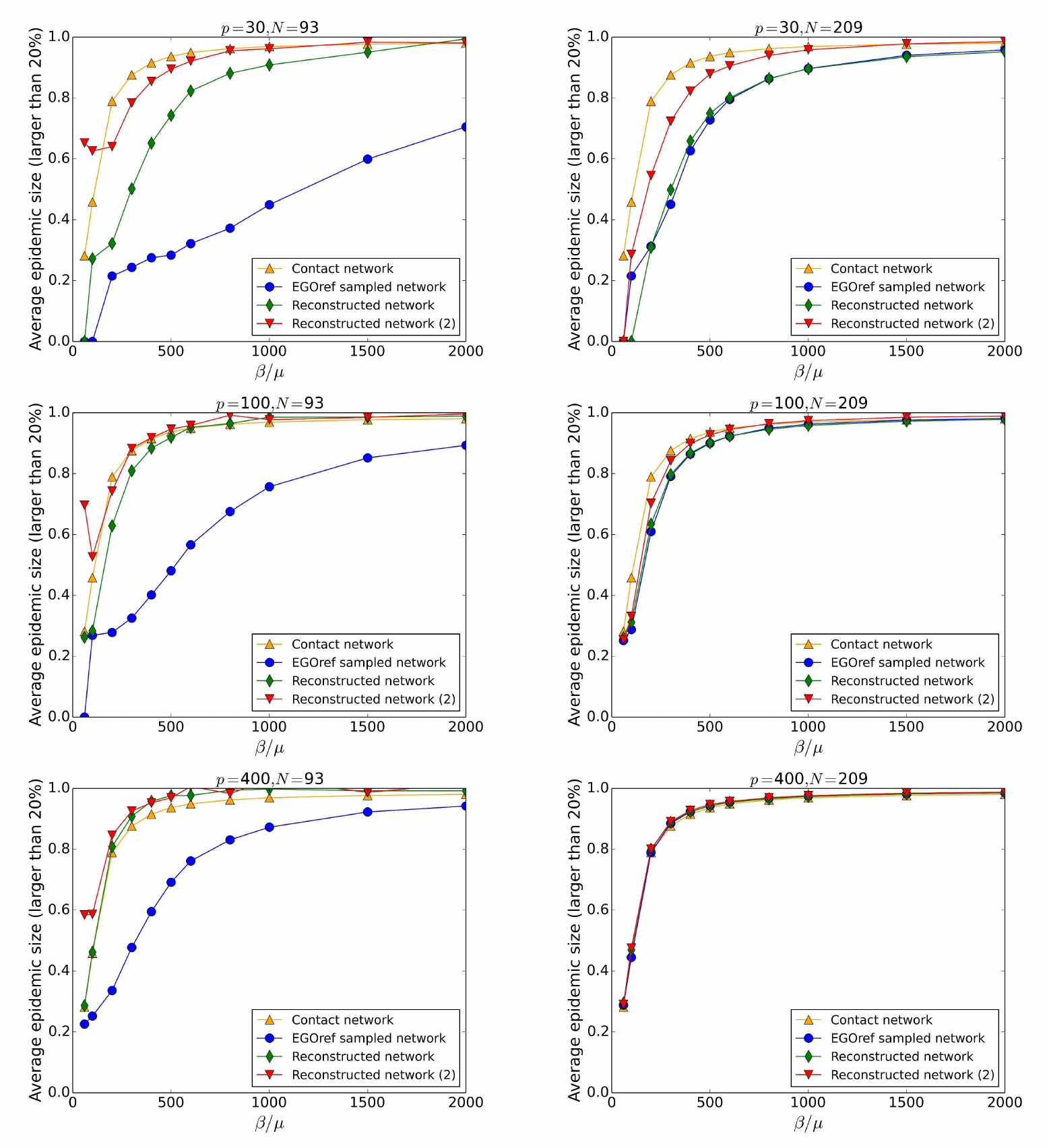}
        \caption{\textbf{InVS dataset: Outcome of SIR spreading
            simulations.} Average size of epidemics with size above
          20\% as a function of  $\beta/\mu$
          for several values of $p$ and $N$. The simulations are
          performed on the contact network on the EGOref sampled network and on the surrogate data obtained
          using two different methods of reconstruction.}
        \label{fig:epi1_invs}
\end{figure}

\begin{figure}[htb]
        \centering
        \includegraphics[width=0.8\textwidth]{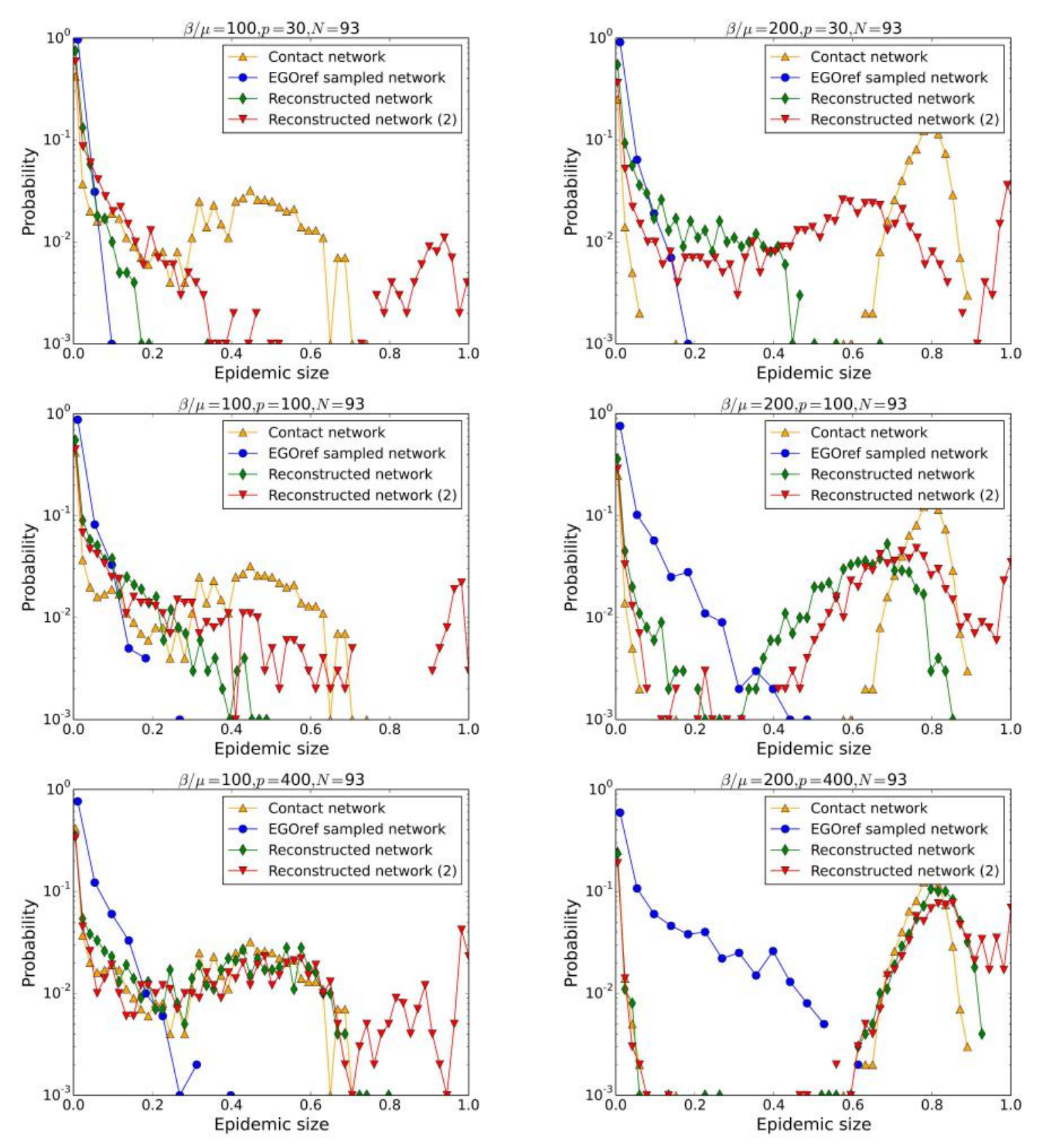}
        \caption{\textbf{InVS dataset: Distributions of epidemic sizes
            of SIR spreading simulations.} The parameter of spreading
          $\beta/\mu$, the parameter $p$ and the number of sampled
          individuals $N$ are given above each figure. The simulations
          are performed on contact network, EGOref sampled network and
          the reconstructed networks using two different methods of
          reconstruction.}
        \label{fig:epi2_invs}
\end{figure}

Figures \ref{fig:epi1_invs} and \ref{fig:epi2_invs} compare the outcomes of SIR simulations performed on the contact network, 
the sampled and the reconstructed networks, for various values of the sampling and spreading parameters.
As for the Thiers13 case, the simulations on the sampled networks strongly underestimate the 
epidemic risk, except obviously at large $p$ and $N$. The use of surrogate data generally 
improves the estimation of the epidemic risk, except at large $N$ for the first reconstruction method, as 
in this case almost no nodes have to be added so the network is almost unchanged by this method.
For small and intermediate $p$, the second method gives better estimations of the average epidemic size while, at 
very large $p$, the effect of assigning randomly weights leads as before to a slight overestimation for reconstructed networks.
At small $\beta/\mu$ and $p$, the second method of reconstruction can even overestimate this size substantially.

Figure \ref{fig:epi2_invs} sheds some more light by displaying the whole distributions of 
epidemic sizes for several values of the spreading parameter 
$\beta/\mu$ and of the sampling parameter $p$, for a rather small $N$ of the order of $40\%$ of the total population. 
While the distributions obtained with the second reconstructed networks are generally closer to the ones obtained with the original
contact data than with the EGOref sampled network (which always leads to a very strong underestimation) or the first reconstruction
method, a peak at very large epidemic sizes is observed for the reconstructed data, which is not present for the original contact data. 
The maximal sizes of epidemics is thus overestimated.
As in the Thiers13 case, this effect is likely to be due to the 
fact that the amount of weights on inter-departments edges is larger in the surrogate data than in the original data, 
allowing for a more efficient spread across the whole population.

\section*{Discussion}

This paper positions itself in the context of the issue of data
incompleteness in contact networks. More specifically, since many
datasets are de facto incomplete, it is important to assess how data
incompleteness affects the outcome of data-driven simulations, how the
resulting biases can be compensated, and how much data is needed for
the simulations
\cite{Genois:2015,Mastrandrea:2016,Fournet:2016,Stehle:2011,Machens:2013}.
We have here considered the case of non-uniformly sampled contact data
and focused on a sampling procedure designed to mimic data resulting
from surveys or diaries \cite{Fournet:2016}. This sampling procedure
results in both population sampling, as not all individuals in the
population are respondents, and in non-uniform link sampling, to mimic
the fact that longer contacts have a larger probability to be
remembered or to correspond also to friendship links.

We have applied this sampling procedure, called EGOref, on two
datasets of contact networks in two different contexts and varied its
two parameters, which determine the population participation rate and
the fraction of sampled links. The datasets concern populations
structured in groups with very different mixing patterns: in a high
school, the class structure strongly determines contacts, with more
than $90\%$ of the duration of contacts occurring within classes; in office
buildings on the other hand, the population is divided into
departments but the impact on the contacts is less strong.  As
expected from previous investigations, using sampled data to run
simulations of spreading processes leads to a strong underestimation
of the epidemic risk, as quantified by the distribution of epidemic
sizes. We have therefore considered the issue of building surrogate data from
the sampled data, such that simulations using the surrogate data yield
a better estimation of the epidemic risk. The first method we
envisioned has been shown to yield good results in the context of
uniform population sampling \cite{Genois:2015}. It is based on the
fact that the uniform sampling keeps invariant the contact matrix
giving the densities of links between groups in the population. We
have shown that the resulting surrogate data, when built from sampled data
using the EGOref procedure, yields better estimations than the raw
sampled data, but still yields a largely underestimated risk, since
the link sampling of the EGOref procedure leads to a sampled network
with a (possibly much) lower density than the original one, and the
method of Ref. \cite{Genois:2015} does not compensate for this bias.
This implies that more information is needed on the data than just the
contact matrix of densities measured in the sampled data.  One of the
simplest way to add information to the sampled data is to assume that
the link density of one of the groups of the population is known: this
can occur for instance if a measurement of contacts using wearable
sensors is feasible only for a small subset of the population, for
practical reasons, so that it is possible to correctly measure the
link density for that group. We therefore considered a second method of
construction of surrogate data, in which we first rescaled all the
elements of the contact matrix measured in the sampled data by the
ratio between the known and measured densities of the "known" group.
The resulting surrogate contact network has thus a density closer to
the original one.

In the case of a strongly structured population, such as the high
school, we have obtained a strong improvement of the results and a good
estimation of the epidemic risk, for a large range of realistic
parameters, even if the maximum size of epidemics is slightly
overestimated. We have also shown that this method gives good results when
applied on the data obtained from a survey asking students about their
friendship relations. In the case of the less-structured population
(offices), we obtained a clear improvement of the epidemic risk
estimation, but the probability of very large epidemics becomes
strongly overestimated when using the surrogate data. We have linked
this overestimation with the fact that the total amount of weights
carried by the inter-class or inter-department edges is larger in the
surrogate than in the original data. Indeed, as shown in
Ref. \cite{Genois:2015}, intra-class and intra-department links tend
to carry larger weights than inter-class and inter-department ones. As weights are distributed randomly
in the surrogate data, without taking this into account, the reconstruction tends
to attribute more weight to the inter-groups edges with respect to the original data, which 
favors the spread. 

Our results are overall two-sided. On the one hand, it is remarkable that, using very little information,
namely the contact matrix of the sampled data and if possible the knowledge of the density of one single group,
using surrogate data instead of the raw sampled data
leads to a strong improvement of the epidemic risk estimation as quantified by the fraction of large epidemics
and the average size of these epidemics. On the other hand, 
the lack of information about the precise values of the densities of links between pairs of groups,
about the relative weights of intra- and inter-groups edges,
as well as about the potential existence of small cohesive substructure, may lead, when the surrogate data is used, to 
distributions of epidemic sizes differing from the original ones, with for instance the over-estimation
of the largest epidemic sizes and the presence of a peak at very large epidemic sizes.
This shows both that survey data can be effectively used to construct surrogate data, but also
that more detailed information coming from data collection with wearable sensors is of enormous value, even if such
collection concerns only a fraction of the population, as it allows
(i) to have a correct estimation of the overall density, (ii) to obtain the distribution of 
contact durations, (iii) if enough sensors are available, to obtain a much better picture of the contact matrix and
of the fraction of weights corresponding to intra- and inter-groups contacts, all elements having a role in the
unfolding of spreading processes in a population.

Our study contributes to the discussion on the amount of details
actually needed in contact data to be used in data-driven models. It
is worth noting some of its limitations. We have focused here on one
specific sampling model, while others might be of interest. We argue
that this procedure is particularly relevant as it mimics surveys or
diaries, as described in Ref. \cite{Fournet:2016}. The procedure also
assumes a uniform node sampling, while positive or negative
correlations with actual contact activity might exist. Additional
investigations concerning such non-uniform population sampling would
certainly be of interest, as well as studies on other datasets or on
synthetic populations with tuneable characteristics. Finally, the effect of using sampled or surrogate
data for data-driven simulations of other types of processes ought to be investigated. Preliminary simulations of
the Susceptible-Infectious-Susceptible model (SIS), in which individuals who recover become again susceptible and can catch again the disease,
show that results similar to the SIR case are obtained (not shown): the sampling leads to a strong underestimation of the epidemic risk that is 
compensated only partially by the first method of reconstruction; the second method gives excellent results for the very structured dataset
but still leads to an underestimation of the epidemic risk for the offices dataset. The issue of the performance
of using surrogate data in more complex processes such as complex contagion remains open for future investigations.

\section*{Author contributions statement}
A.B. and J.F. conceived and designed the study. J.F. performed the numerical simulations and the statistical analysis.
A.B. and J.F. wrote the manuscript.

\section*{Competing financial interests}
 The authors declare no competing financial interests.

\newpage
\clearpage

\setcounter{figure}{0}
\renewcommand{\thefigure}{S\arabic{figure}}

\setcounter{table}{0}
\renewcommand{\thetable}{S\arabic{table}}

\renewcommand{\thesection}{S\arabic{section}}

\begin{center}
{\Large
\textbf{Estimating the epidemic risk using non-uniformly sampled contact data:
Supplementary Information}
}\\
\end{center}

\begin{figure}[h]
	\includegraphics[width=0.45\textwidth]{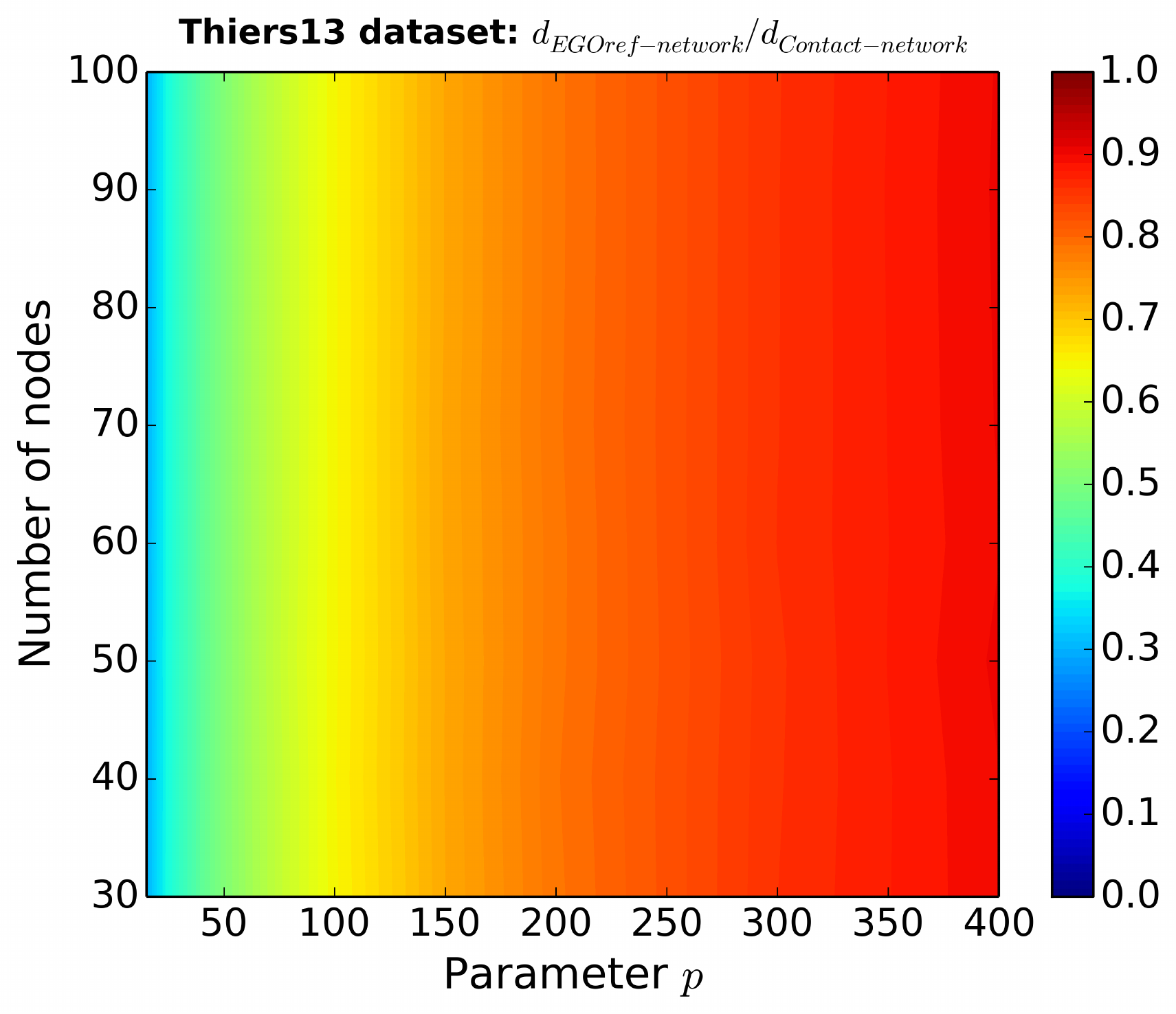}
        \includegraphics[width=0.45\textwidth]{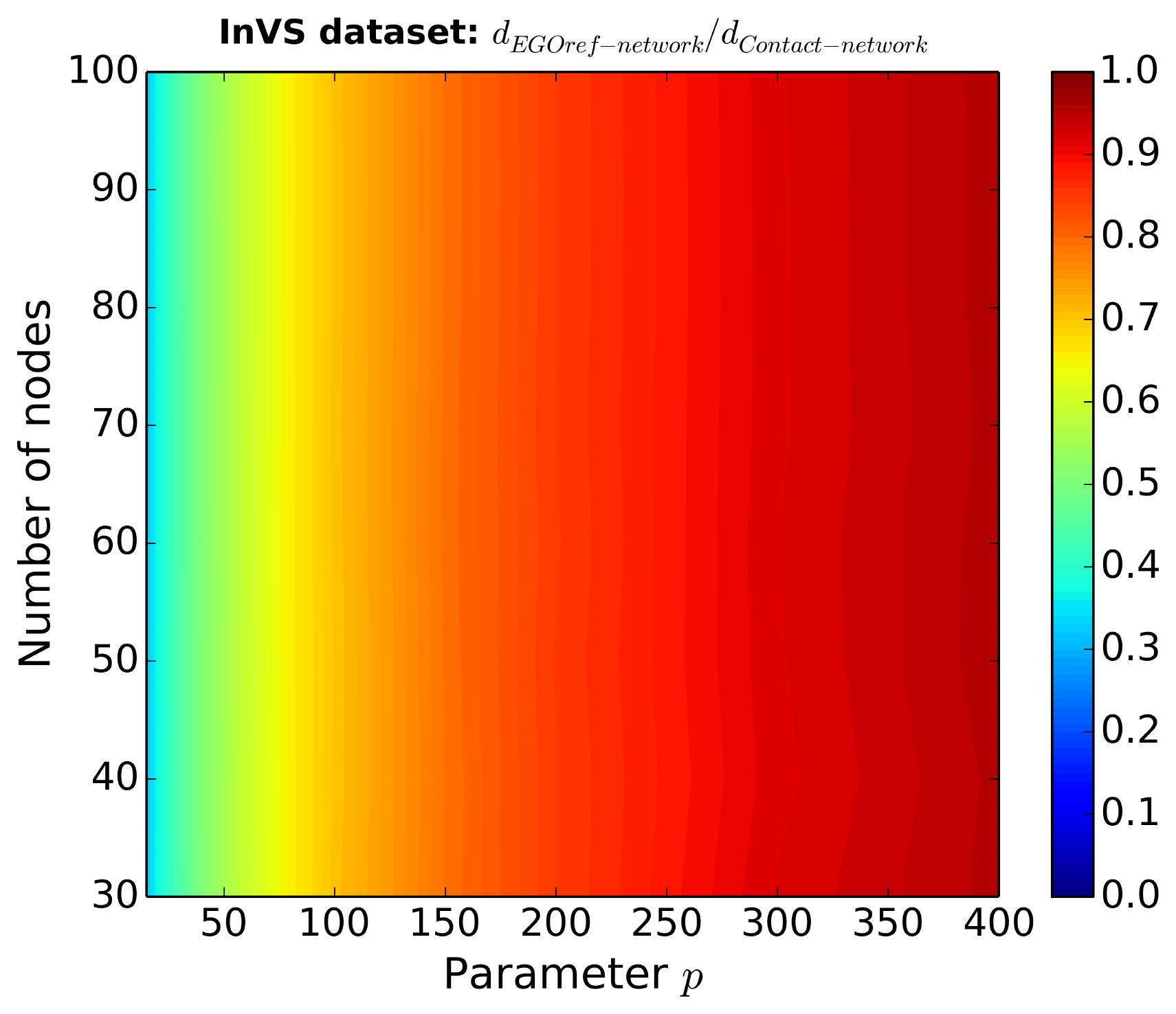}
	\caption{Ratio between the density of the EGOref sampled
          network and the density of the whole contact network as a
          function of the parameter $p$ and of the percentage of sampled
          nodes for the Thiers13 dataset (left) and the InVS
          dataset (right).}
	\label{fig:density_color}
\end{figure}

\begin{figure}[h]
        \includegraphics[width=0.45\textwidth]{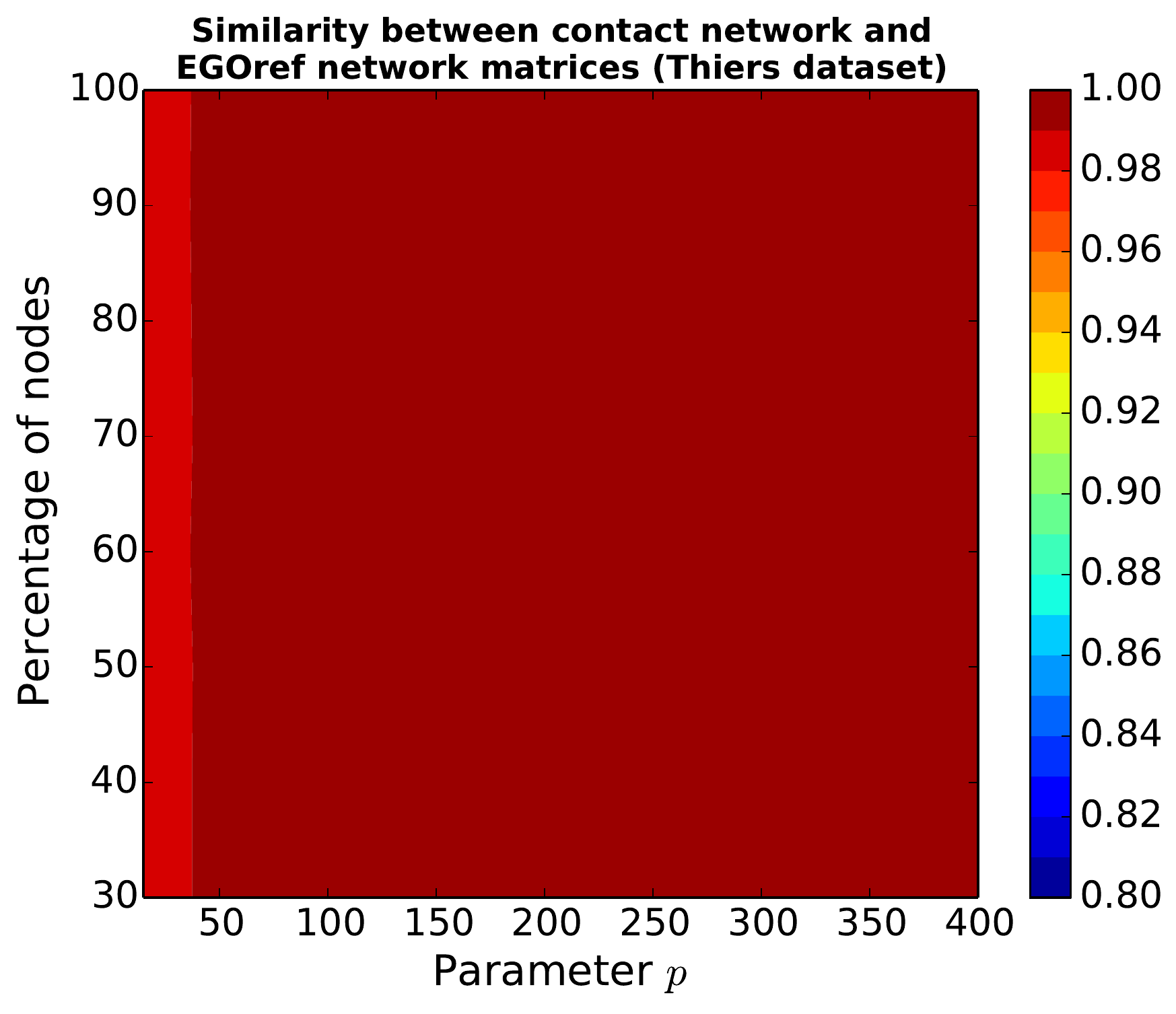}
        \includegraphics[width=0.45\textwidth]{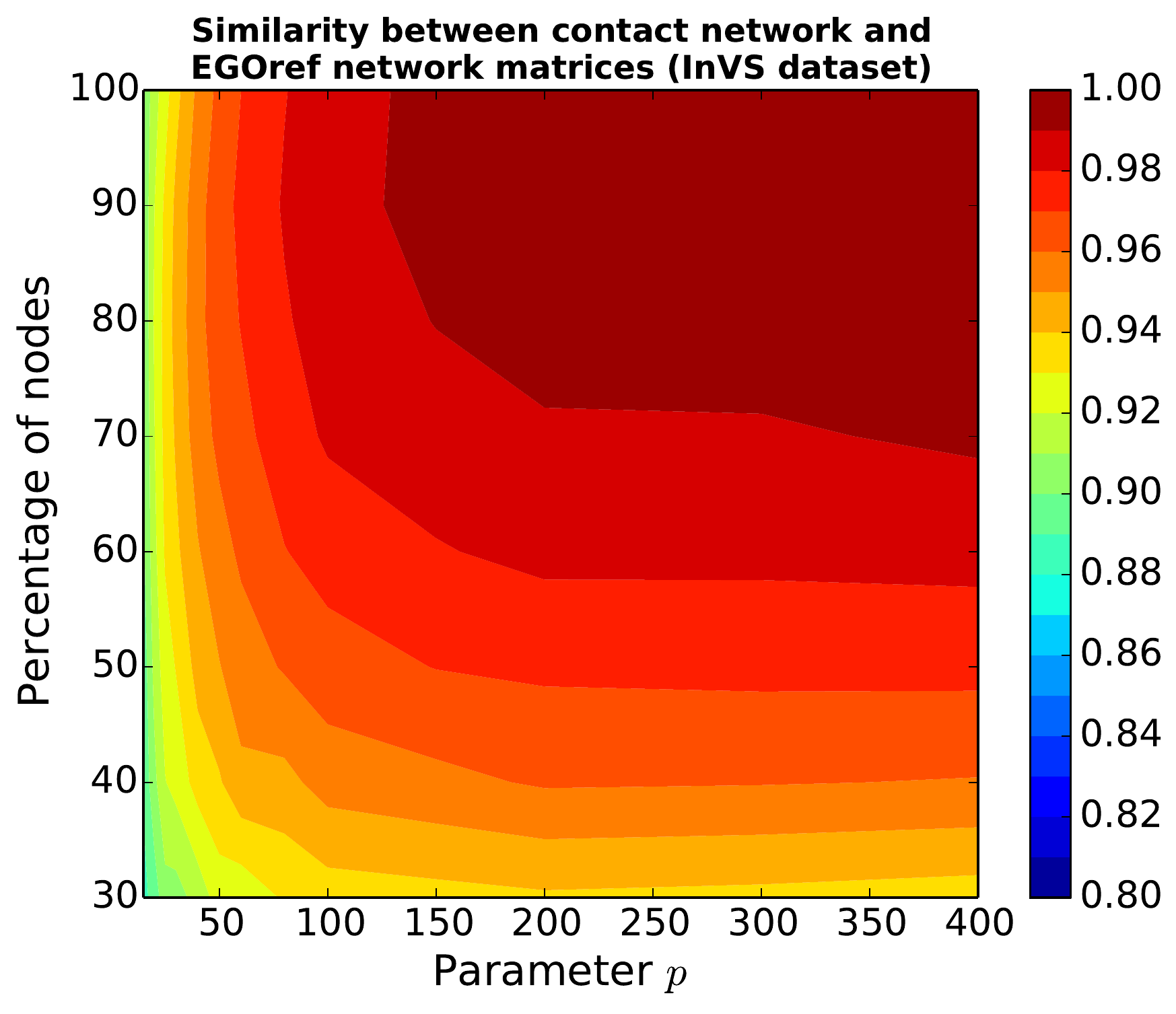}
\caption{Similarity between the contact matrices of the sampled and original networks,  as a
function of the parameter $p$ and of the percentage of sampled
nodes for the Thiers13 dataset (left) and the InVS
dataset (right).}
\end{figure}

\begin{figure}[h]
        \includegraphics[width=0.3\textwidth]{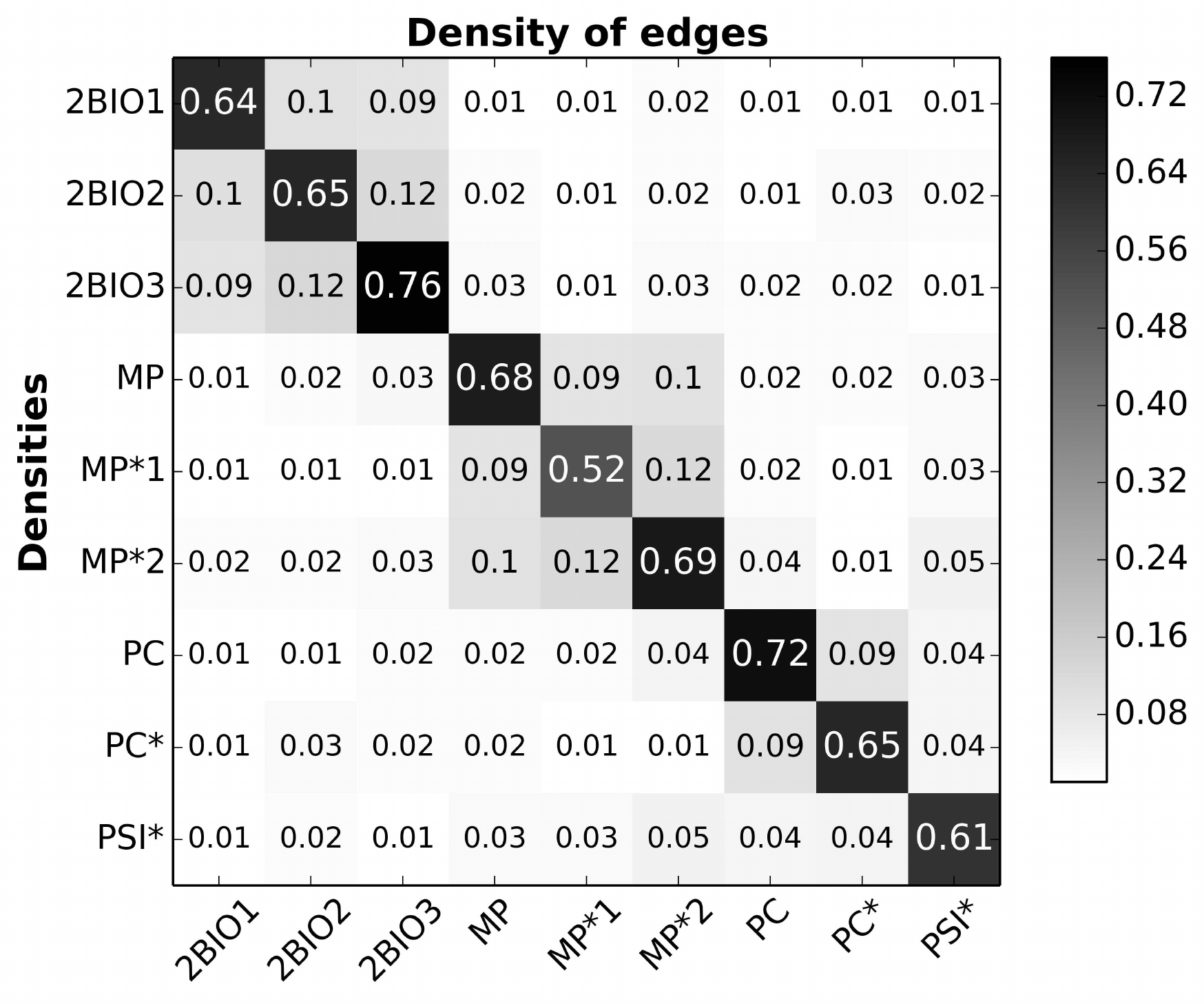}
\includegraphics[width=0.3\textwidth]{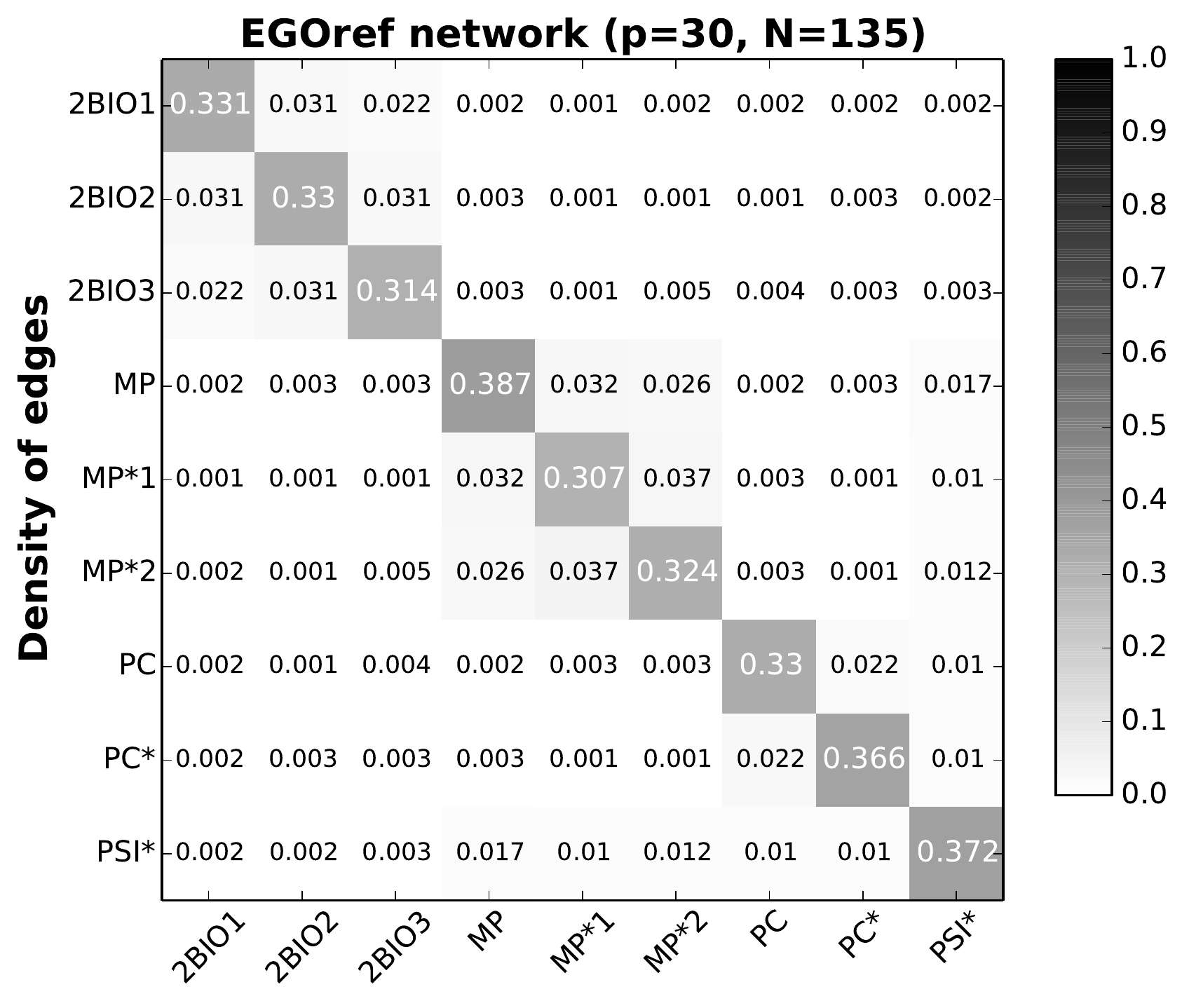}
\includegraphics[width=0.3\textwidth]{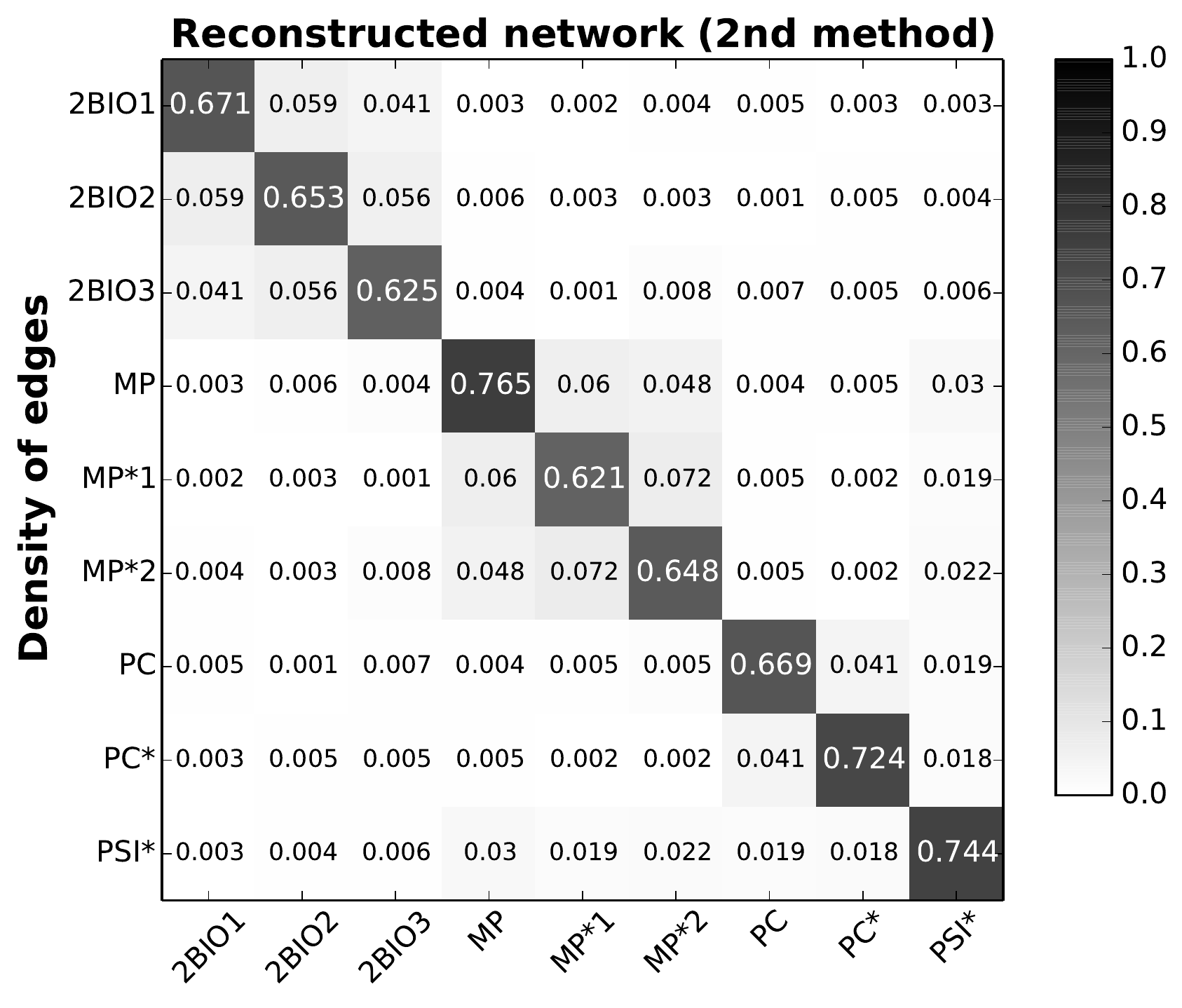}
        \caption{Thiers13 dataset: Contact matrices giving the density of
          edges between departments for the contact network, the
          EGOref network (with $p=30$ and $N$=40\% of the total number
          of nodes) and the reconstructed network using the second
          method of reconstruction. The similarities between the three matrices are all above $98\%$.
 }
        \label{fig:mat_thiers}
\end{figure}

\begin{figure}[h]
	\includegraphics[width=0.3\textwidth]{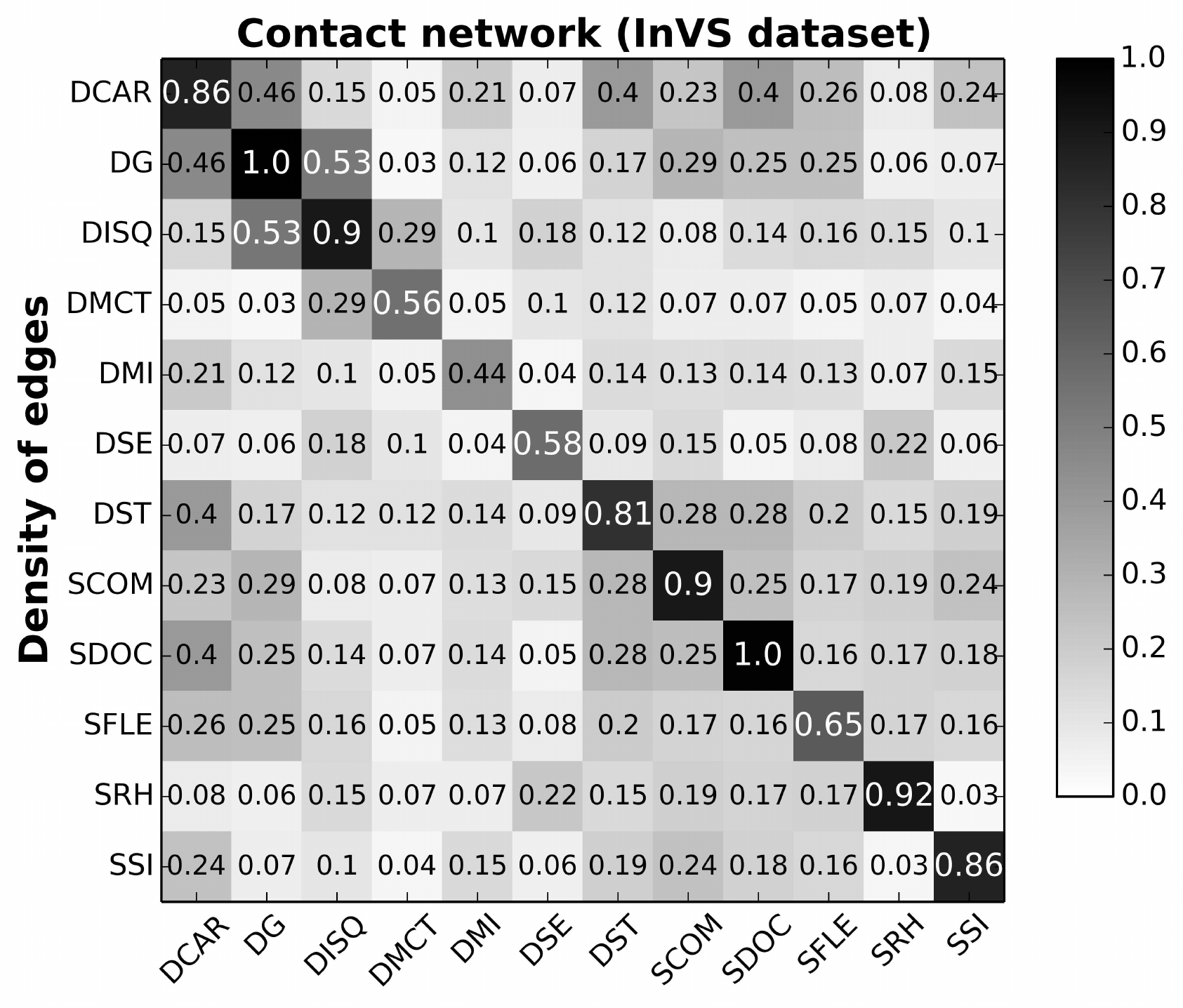}
\includegraphics[width=0.3\textwidth]{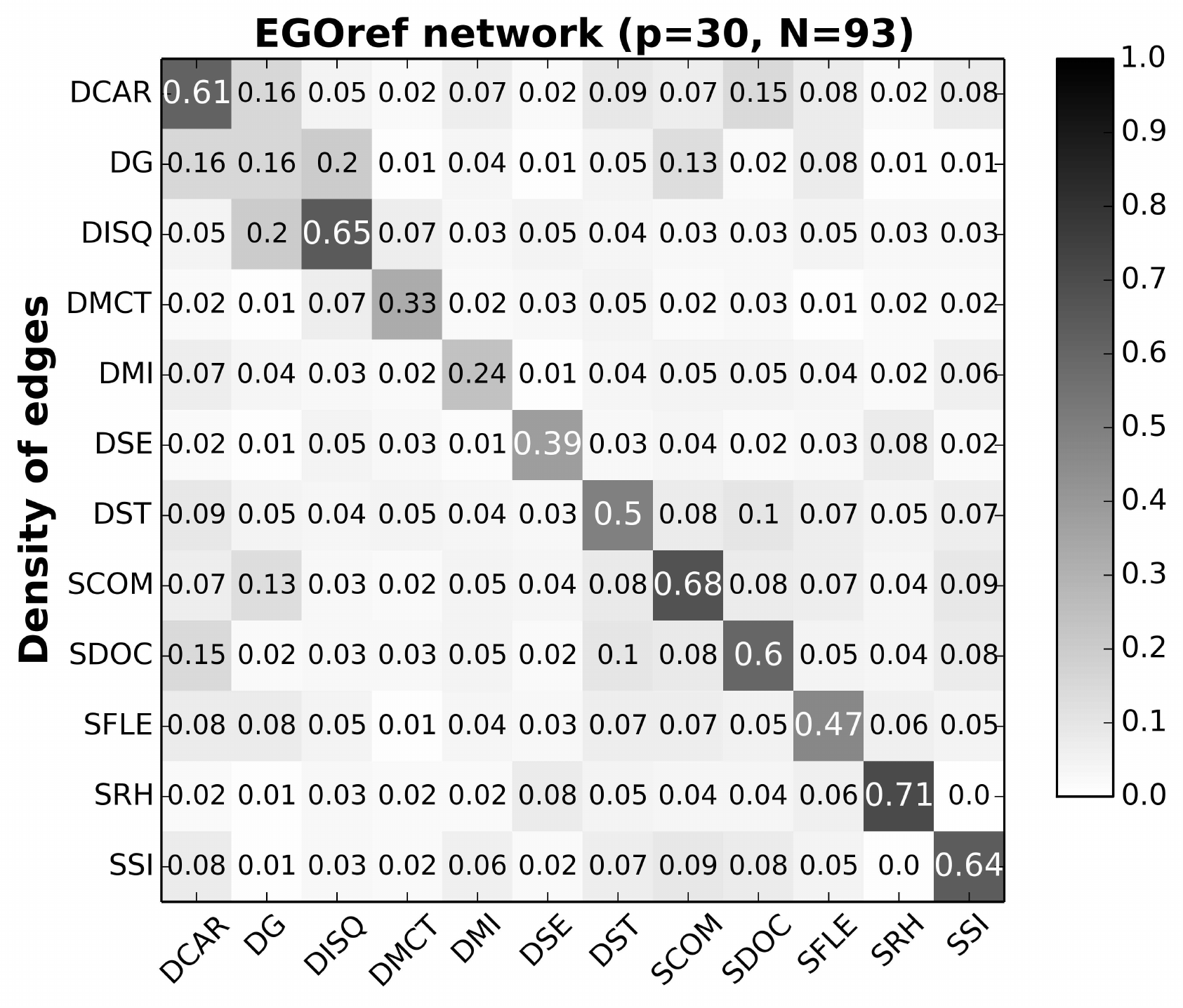}
\includegraphics[width=0.3\textwidth]{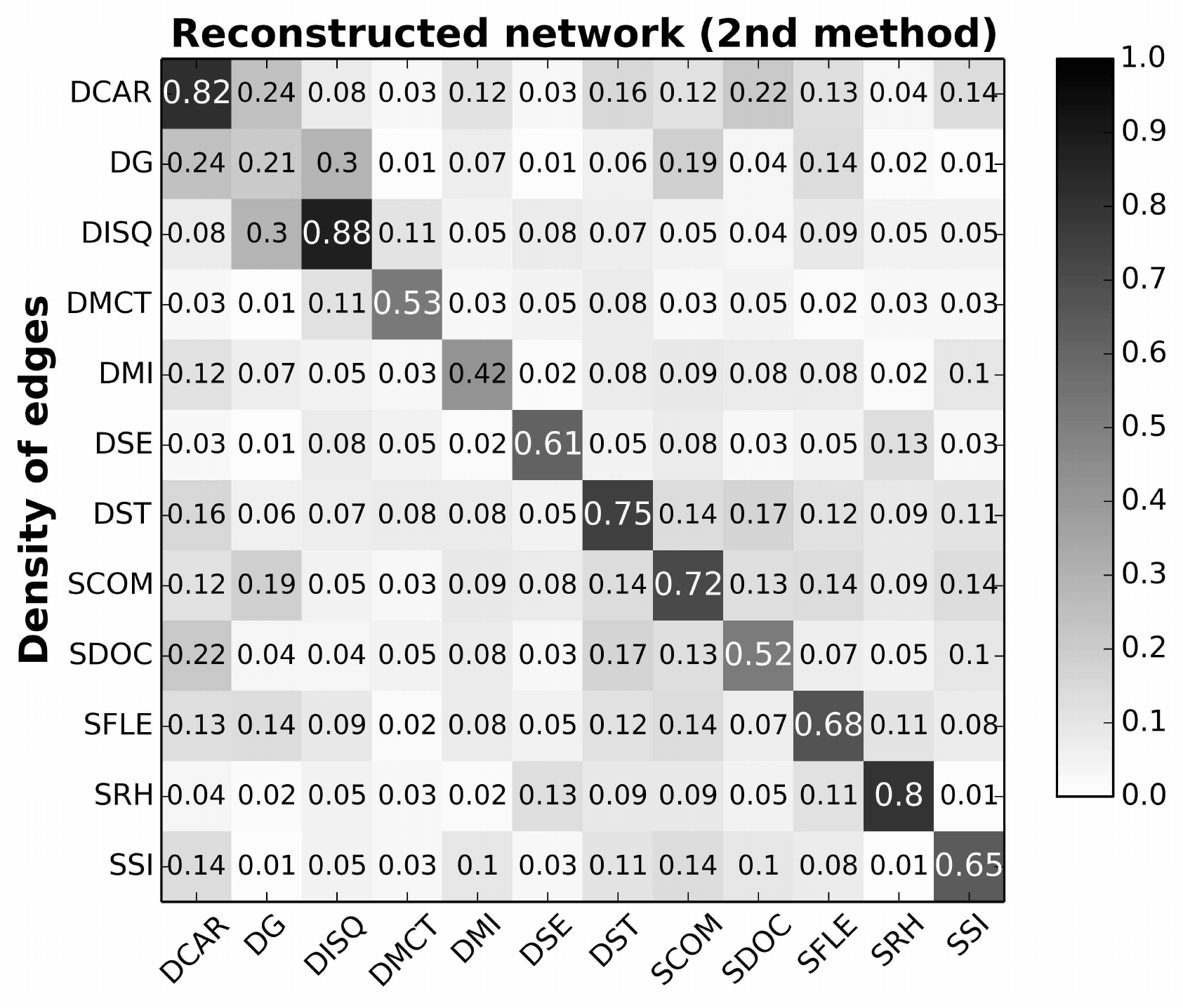}
	\caption{InVS dataset: Contact matrices giving the density of
          edges between departments for the contact network, the
          EGOref network (with $p=30$ and $N$=40\% of the total number
          of nodes) and the reconstructed network using the second
          method of reconstruction. Similarity between the matrix of
          the contact network and of the EGOref network: $93\%$, between
          the matrix of the contact network and of the reconstructed
          network: $98\%$, between the matrix of the EGOref network and
          of the reconstructed network: $94\%$. }
	\label{fig:mat_invs}
\end{figure}

\begin{figure}[h]
        \includegraphics[width=0.45\textwidth]{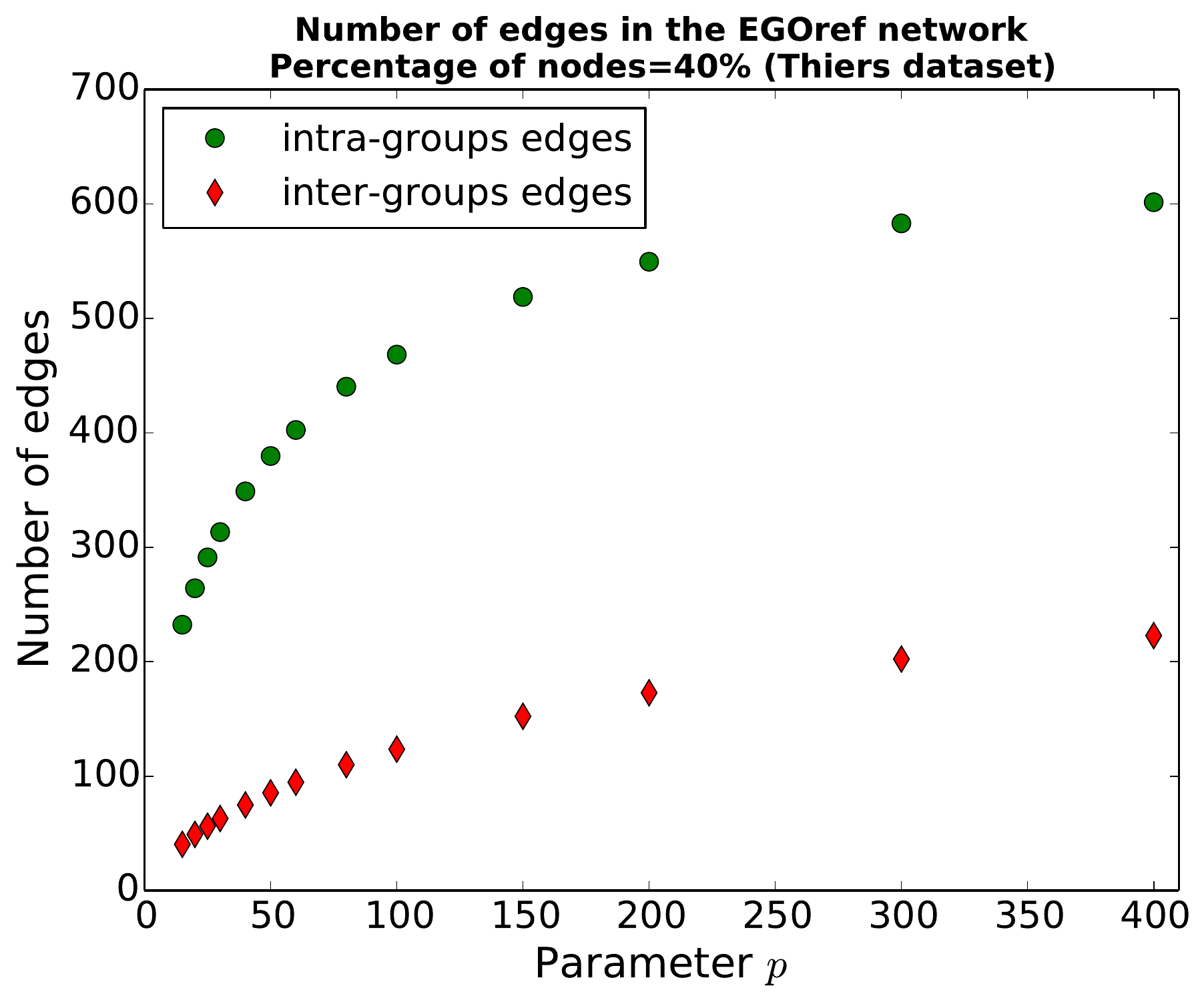}
        \includegraphics[width=0.45\textwidth]{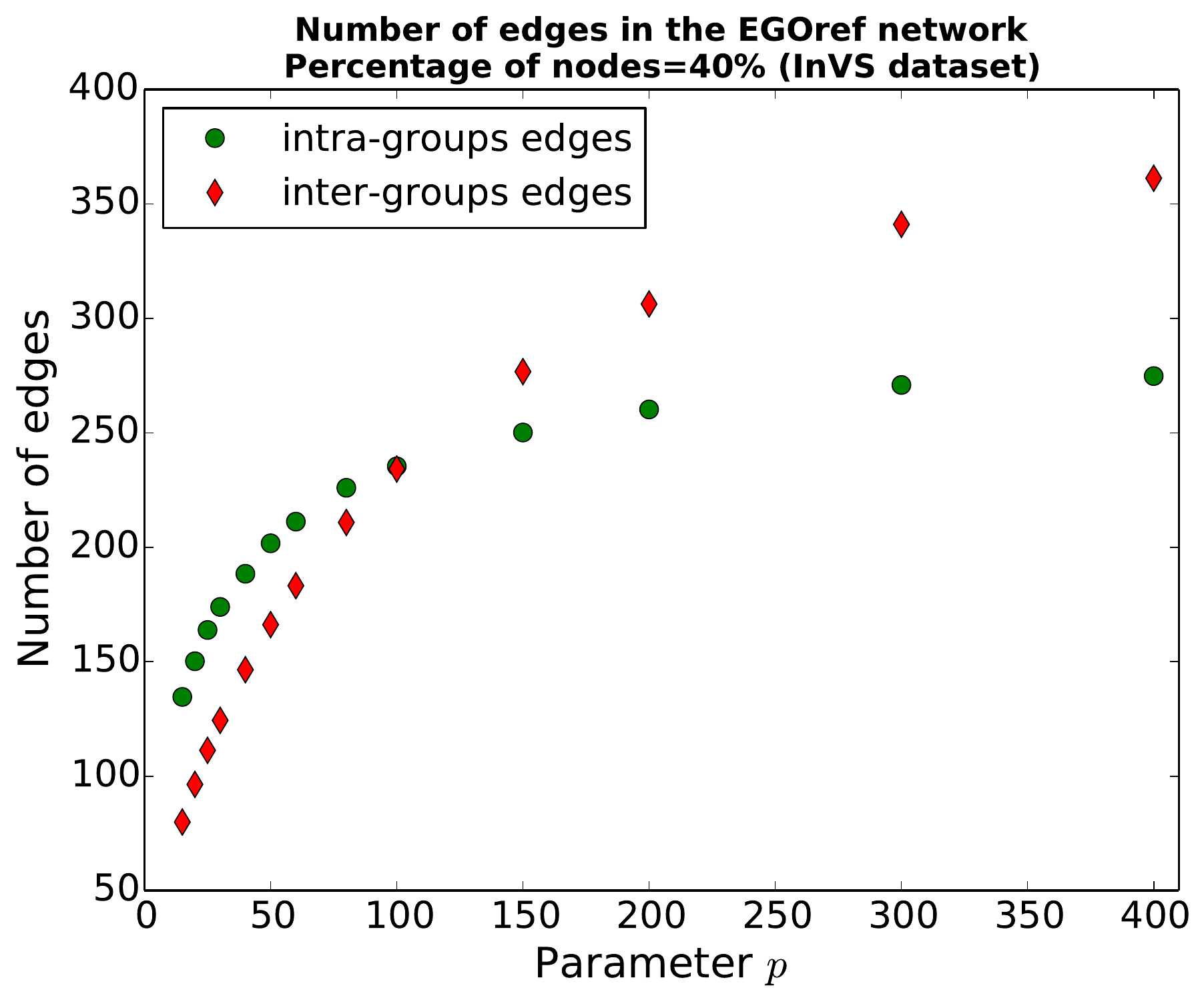}
\caption{Number of intra-class and inter-class edges for Thiers13 and InVS in the sampled network, at varying p and for N = $40\%$.}
\end{figure}

\begin{figure}[h]
        \includegraphics[width=0.45\textwidth]{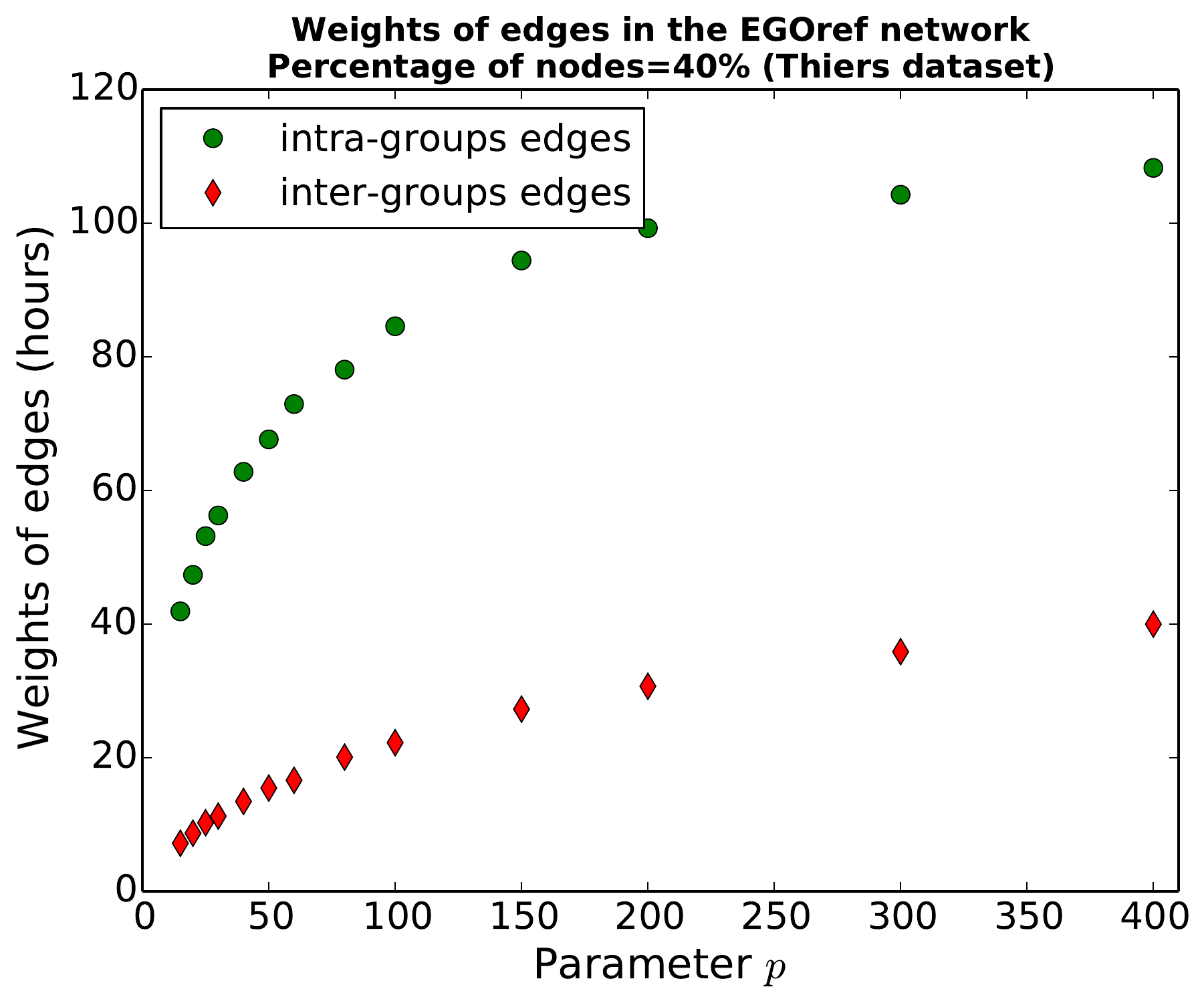}
        \includegraphics[width=0.45\textwidth]{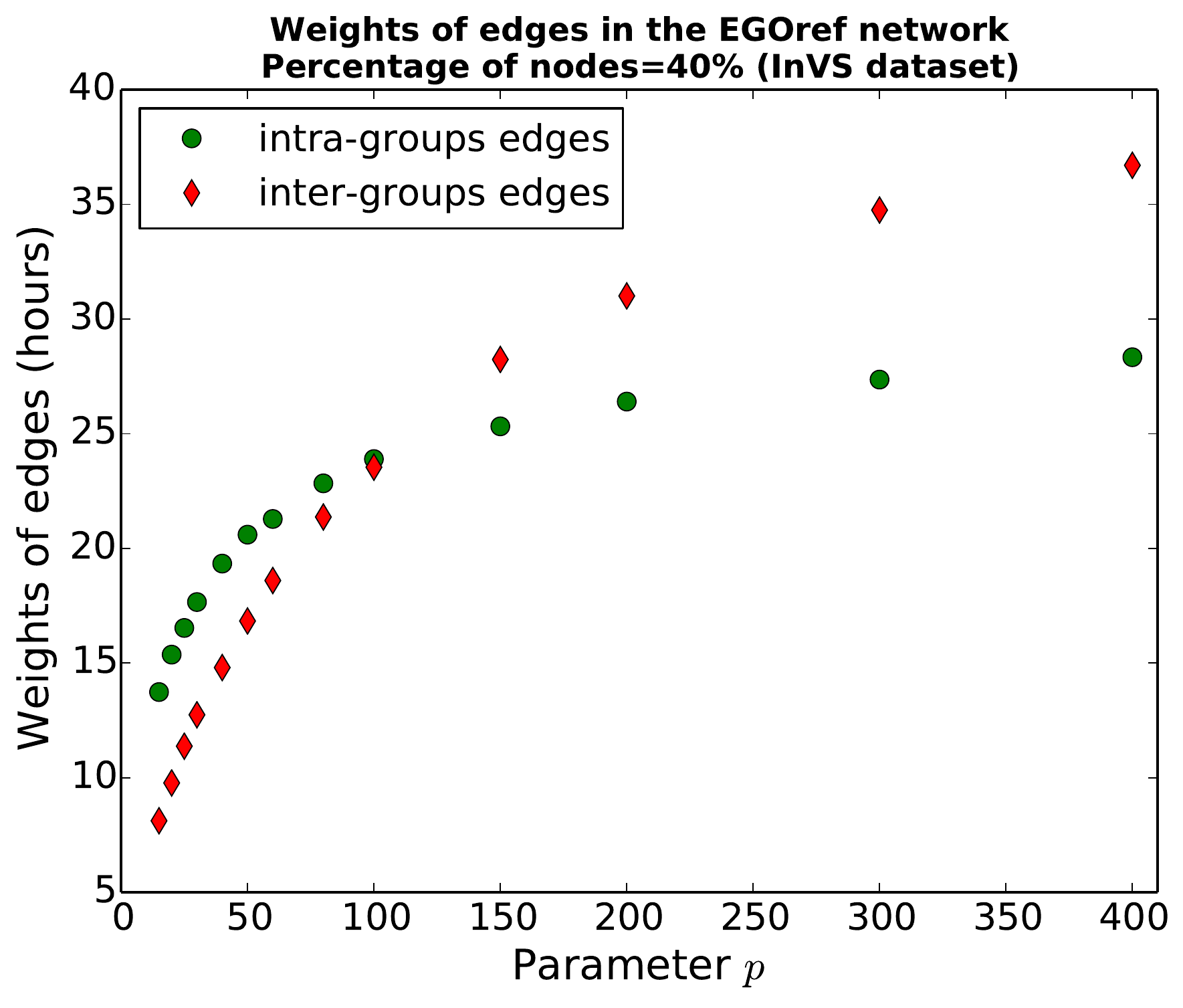}
\caption{Total weight carried by intra-class and inter-class edges for Thiers13 and InVS in the sampled network, at varying p and for N = $40\%$.}
\end{figure}

\begin{figure}[h]
        \includegraphics[width=0.45\textwidth]{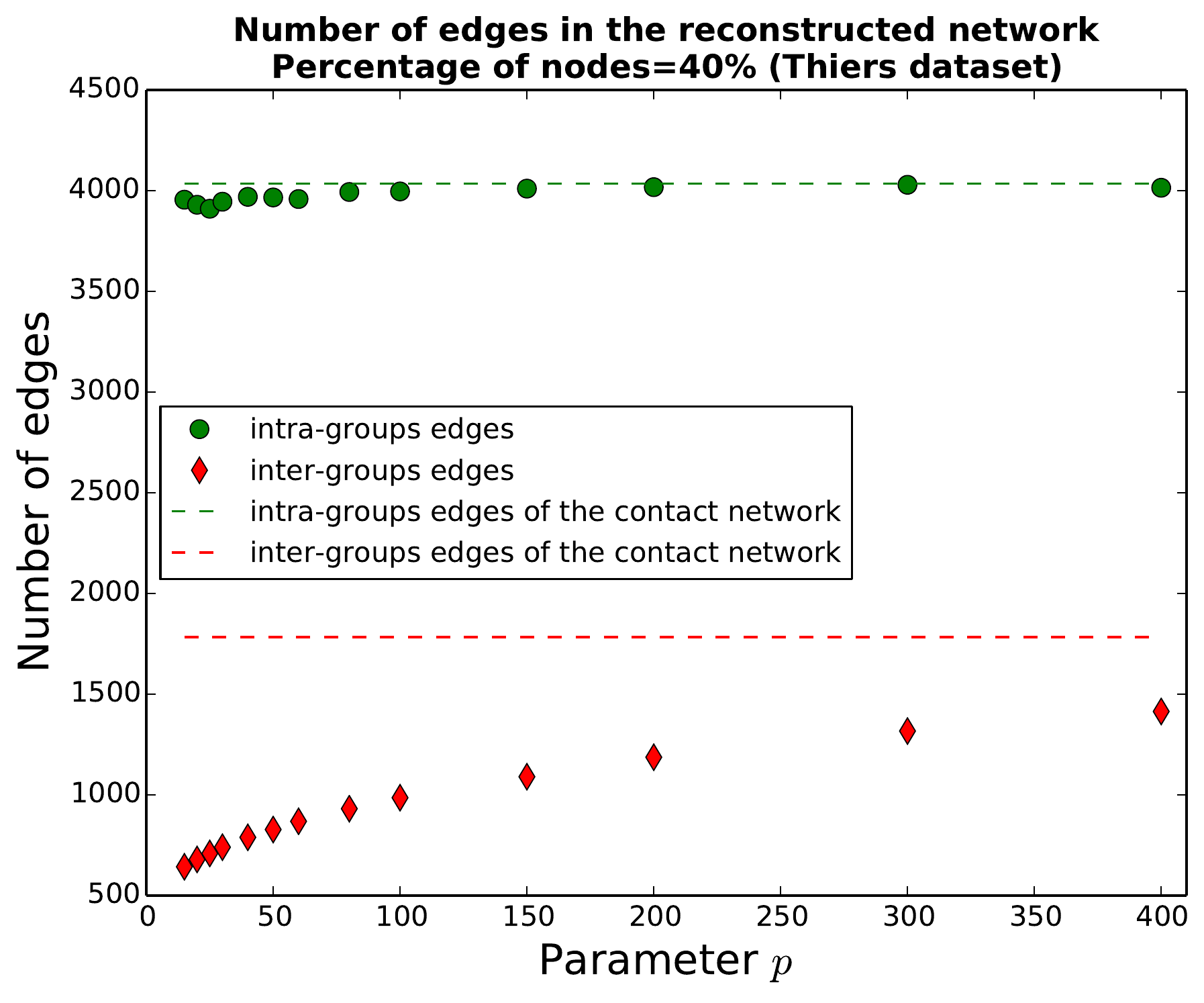}
        \includegraphics[width=0.45\textwidth]{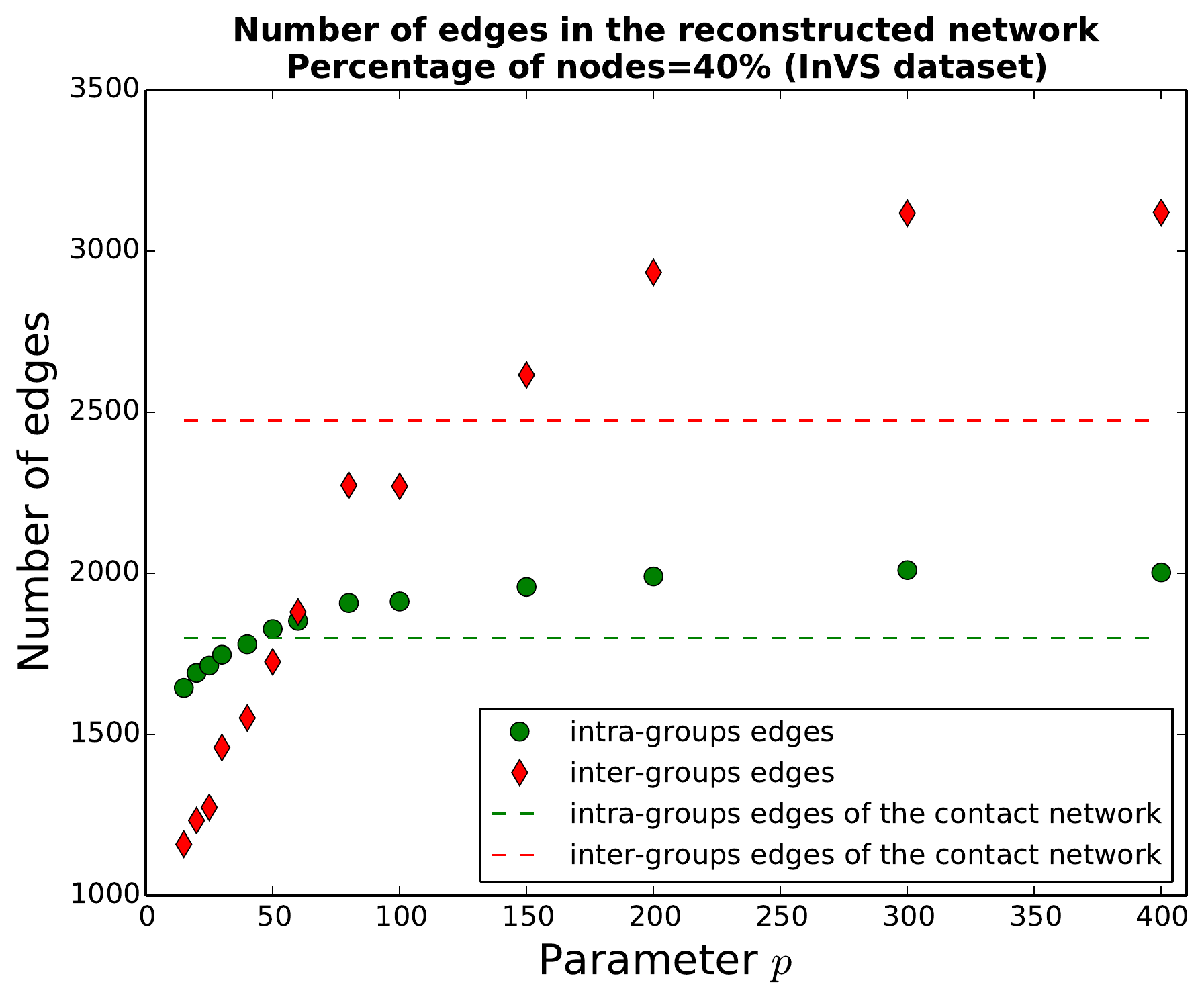}
\caption{Number of intra-class and inter-class edges for Thiers13 and InVS in the reconstructed network, at varying p and for N = $40\%$.
The horizontal lines give the values in the original data.}
\end{figure}

\begin{figure}[h]
        \includegraphics[width=0.45\textwidth]{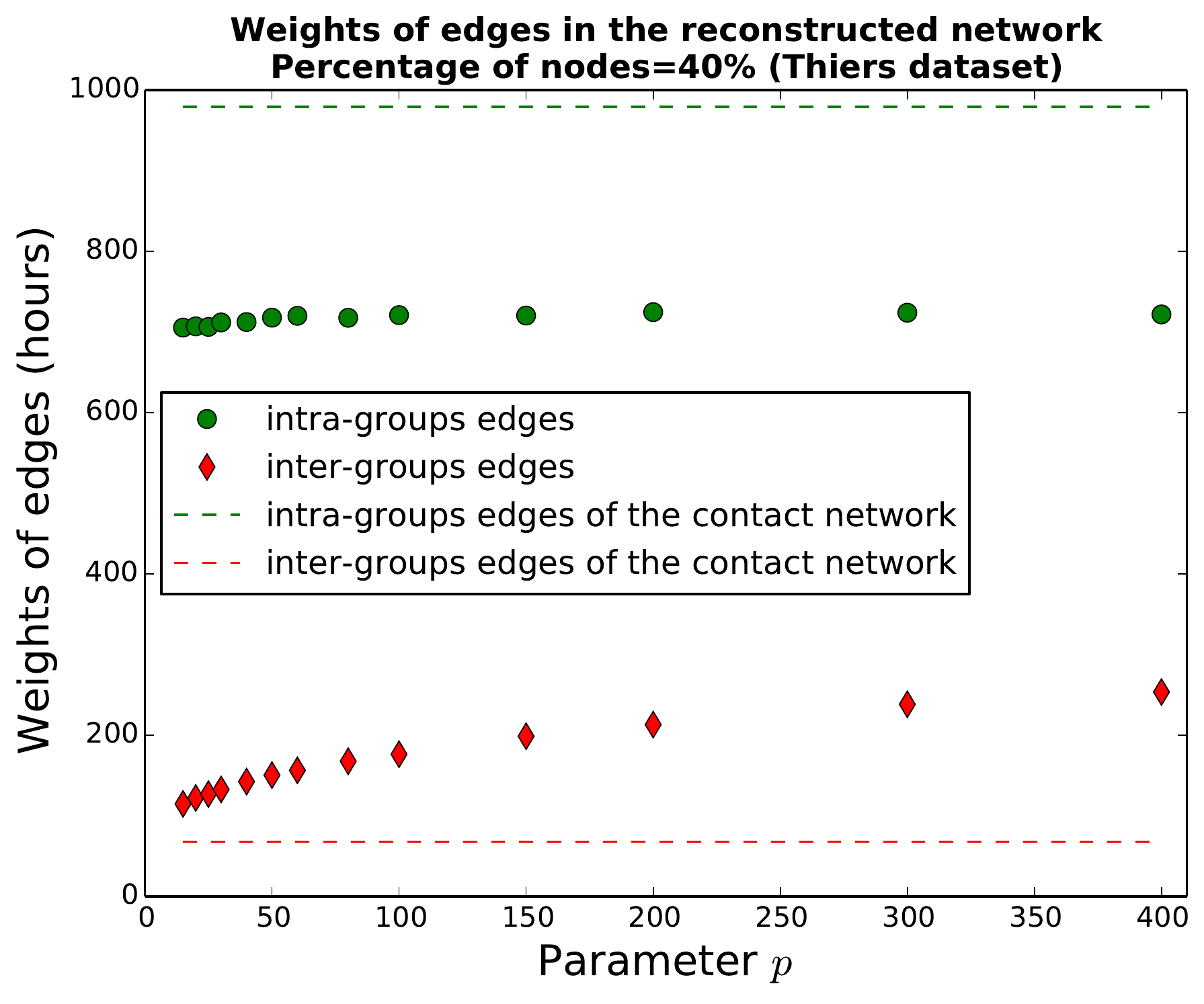}
        \includegraphics[width=0.45\textwidth]{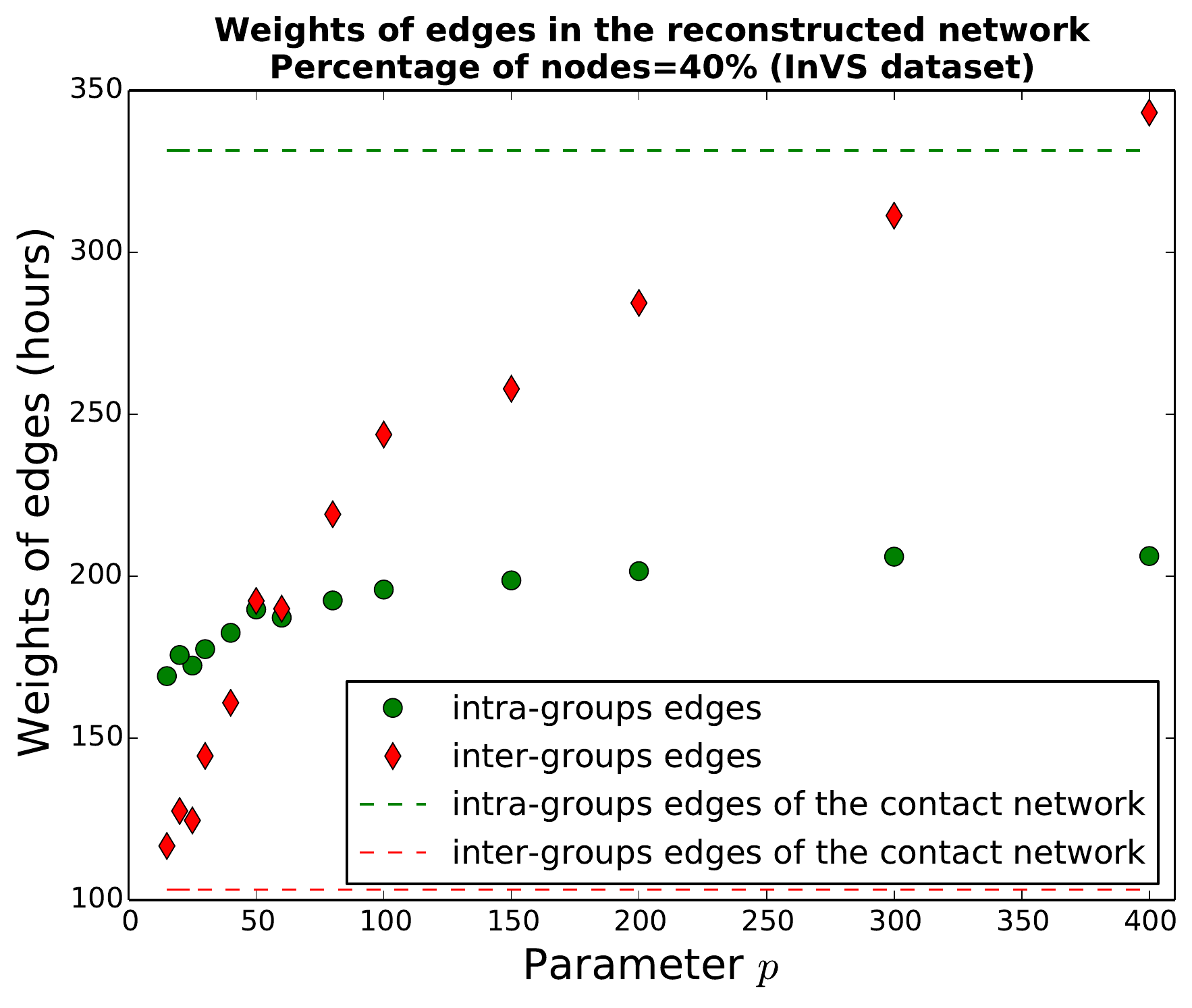}
\caption{Total weight carried by intra-class and inter-class edges for Thiers13 and InVS in the reconstructed network, at varying p and for N = $40\%$.
The horizontal lines give the values in the original data.}
\end{figure}

\begin{figure}[h]
        \includegraphics[width=0.45\textwidth]{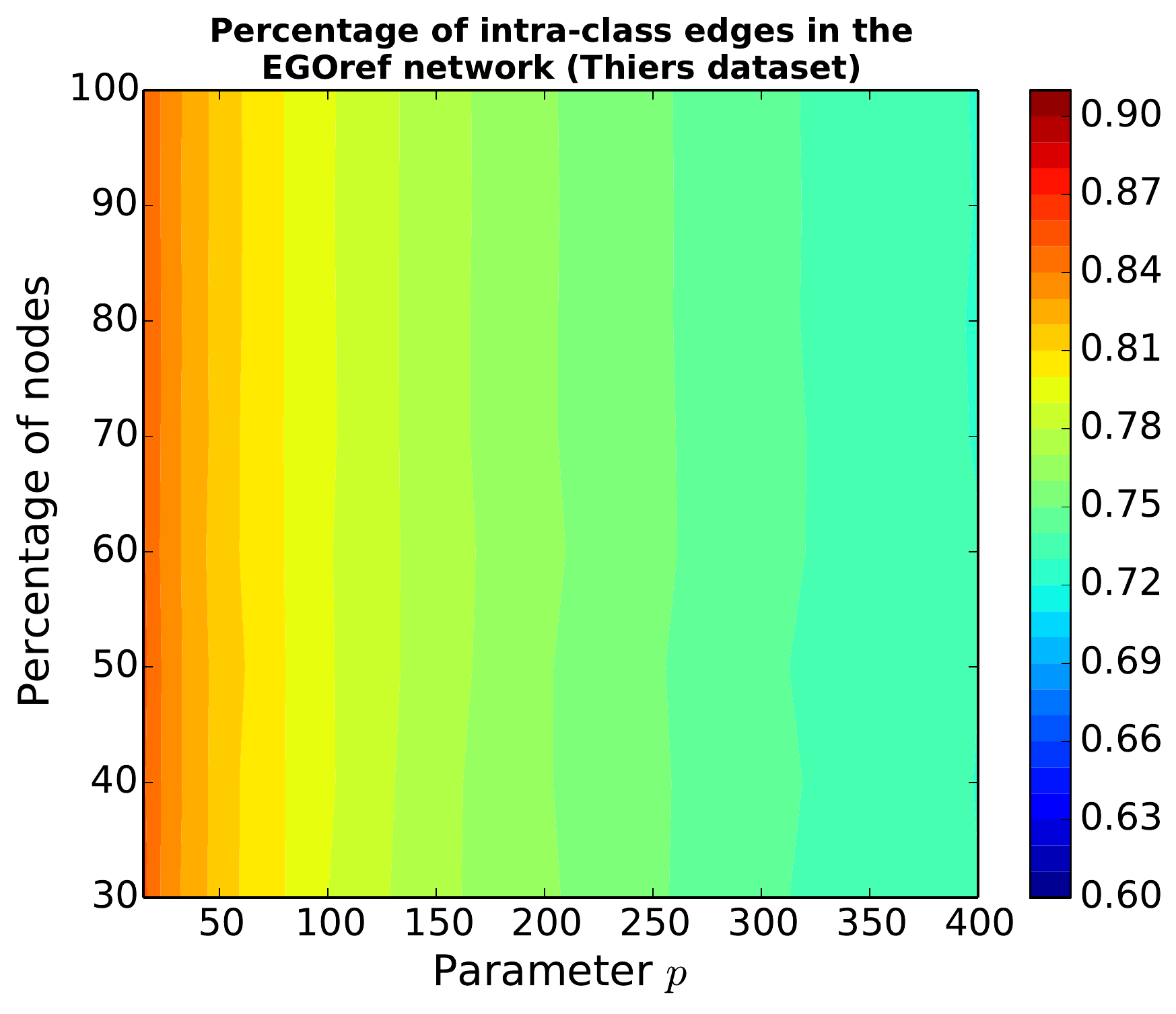}
        \includegraphics[width=0.45\textwidth]{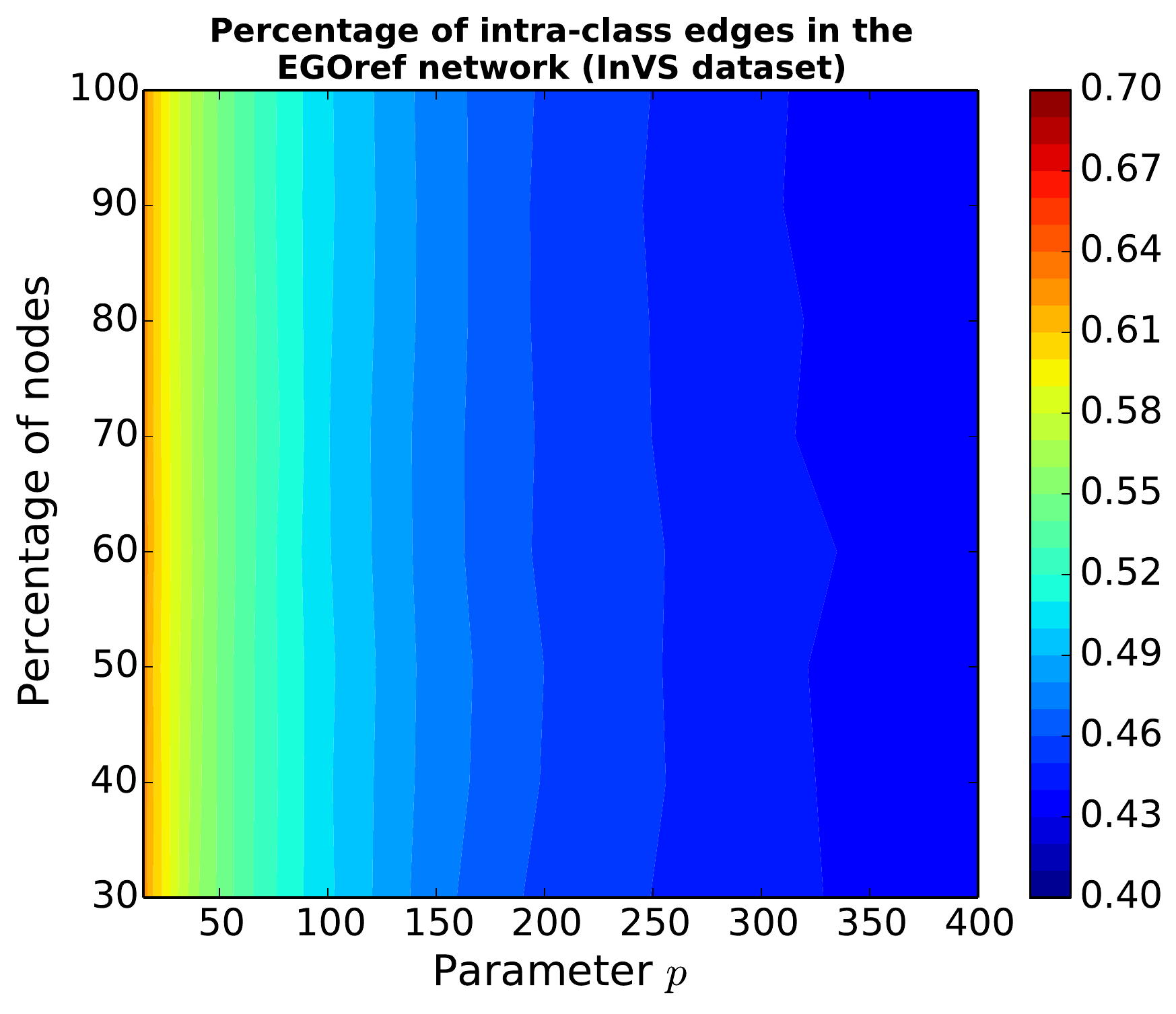}
\caption{Fraction of intra-class edges as a function of the sampling parameter 
$p$ and of the percentage of sampled nodes for the Thiers13 dataset (left) and the InVS
dataset (right).}
\end{figure}

\begin{figure}[h]
        \includegraphics[width=0.45\textwidth]{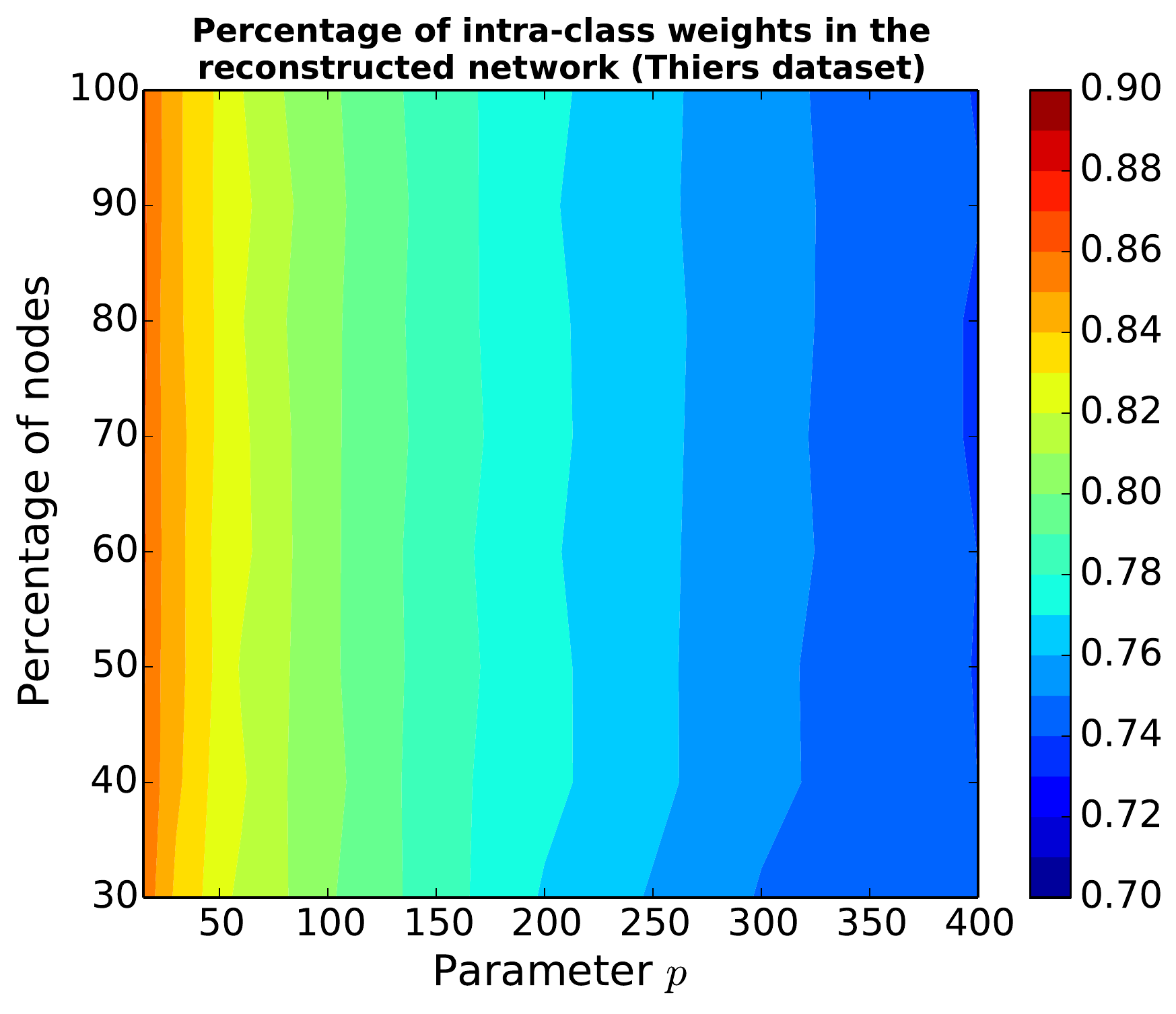}
        \includegraphics[width=0.45\textwidth]{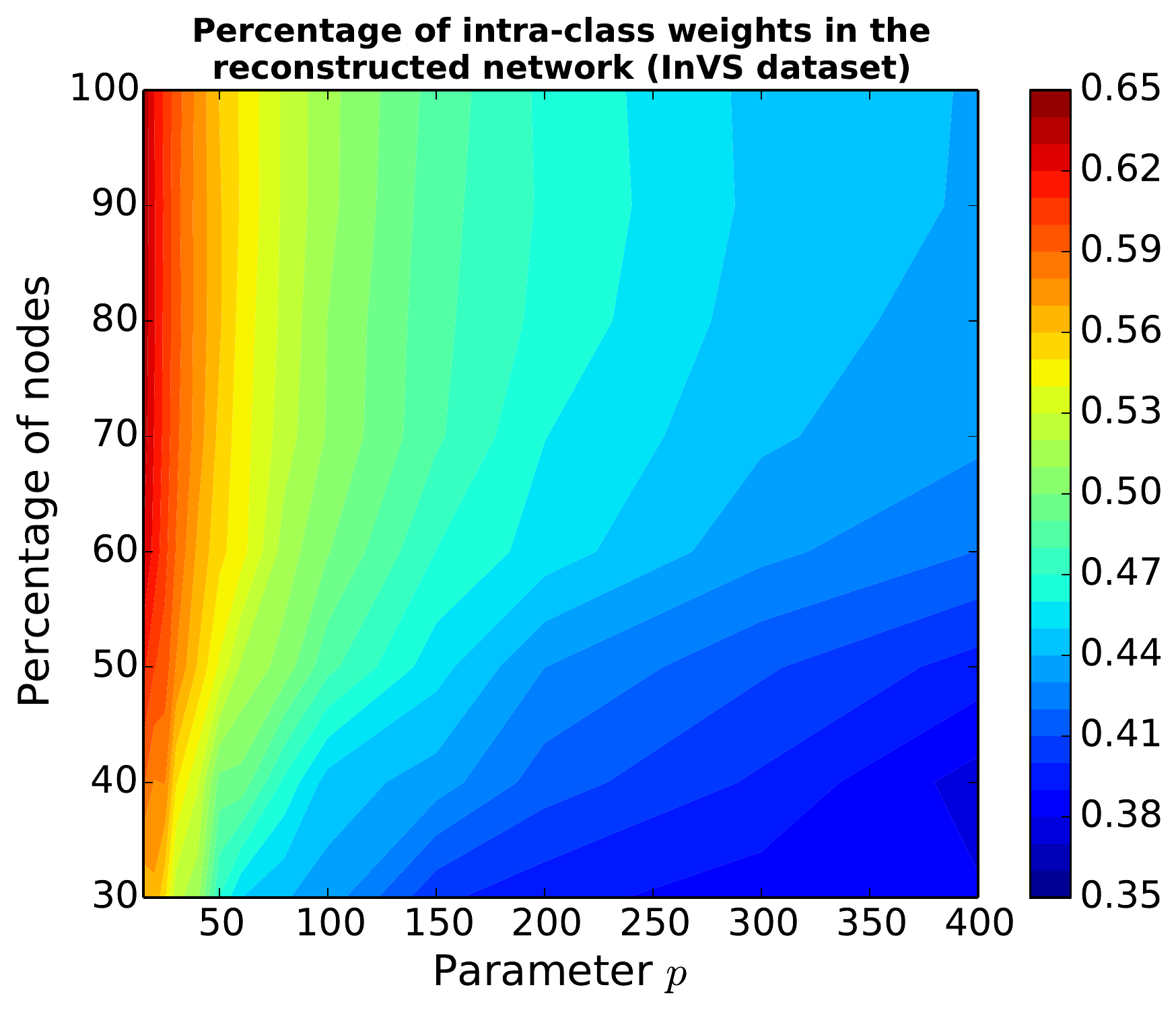}
\caption{Fraction of the total weight carried by intra-class edges in the
surrogate data as a function of the sampling parameter
$p$ and of the percentage of sampled nodes for the Thiers13 dataset (left) and the InVS
dataset (right).}
\end{figure}


\begin{thebibliography}{99}
\expandafter\ifx\csname url\endcsname\relax
  \def\url#1{\texttt{#1}}\fi
\expandafter\ifx\csname urlprefix\endcsname\relax\def\urlprefix{URL }\fi
\providecommand{\bibinfo}[2]{#2}
\providecommand{\eprint}[2][]{\url{#2}}

\bibitem{Eames:2015}
Eames, K., Bansal, S., Frost, S. \& Riley, S. 
Six challenges in measuring contact networks for use in modelling.
{\em Epidemics} {\bf 10,} 72-77 (2015).

\bibitem{Voirin:2015}
Voirin, N. et al.
Combining High-Resolution Contact Data with Virological Data to Investigate Influenza Transmission in a Tertiary Care Hospital.
{\em Infection Control \& Hospital Epidemiology} {\bf 36,} 254 (2015).

\bibitem{Obadia:2015}
Obadia, T. et al.
Detailed Contact Data and the Dissemination of Staphylococcus aureus in Hospitals.
{\em PLoS Computational  Biology} {\bf 11(3),} e1004170 (2015).


\bibitem{Mossong:2008}
Mossong, J. et al. 
Social contacts and mixing patterns relevant to the spread  of infectious diseases. 
{\em PLoS Med} {\bf 5,} e74 (2008)

\bibitem{Mikolajczyk:2008}
Mikolajczyk, R.T., Akmatov, M.K., Rastin, S. \& Kretzschmar, M. 
Social contacts of school children and the transmission of respiratory-spread pathogens. 
{\em Epidemiology and Infection} {\bf 136(6)}, 813-822 (2008).

\bibitem{Pentland:2008}
Pentland, A.
{\em  Honest signals: how they shape our world
(Cambridge, MIT Press, 2008).}

\bibitem{Cattuto:2010} 
Cattuto, C. et al.
Dynamics of person-to-person interactions from distributed RFID sensor networks. 
{\em PLoS ONE} {\bf 5,} e11596 (2010). 

\bibitem{Salathe:2010} 
Salath\'e, M. et al.
A high-resolution human contact network for infectious disease transmission. 
{\em PNAS} {\bf 107,} 22020-22025 (2010).

\bibitem{Conlan:2011}
Conlan, A.J. et al.
Measuring social networks in British primary schools through scientific engagement.
 {\em Proceedings of the Royal Society of London B: Biological Sciences} {\bf 278,} 1467-75 (2011)

\bibitem{Hornbeck:2012}
Hornbeck, T. et al.
Using Sensor Networks to Study the Effect of Peripatetic Healthcare Workers on the Spread of Hospital-Associated Infections.
{\em Journal of Infectious Disease} {\bf 206,} 1549 (2012).

\bibitem{Smieszek:2012}
Smieszek, T., Burri, E.U., Scherzinger, R. \& Scholz, R.W.  
Collecting close-contact social mixing data with contact diaries: reporting errors and biases. 
{\em Epidemiol Infect} {\bf 140,} 744-752 (2012).

\bibitem{Read:2012}
Read, J. M., Edmunds, W. J., Riley, S., Lessler, J. \& Cummings, D. A. T. 
Close encounters of the infectious kind: methods to measure social mixing behaviour.
{\em Epidemiology and Infection} {\bf 140,} 2117-2130 (2012).

\bibitem{Stopczynski:2014}
Stopczynski, A. et al.
Measuring large-scale social networks with high resolution.
{\em PLoS ONE} {\bf 9,} e9597 (2014).

\bibitem{Mastrandrea:2015}
Mastrandrea, R., Fournet, J. \& Barrat, A.
Contact patterns in a high school: a comparison between 
data collected using wearable sensors, contact diaries and friendship surveys.
{\em PLoS ONE} {\bf 10,} e136497 (2015).

\bibitem{Barrat:2015}
Barrat, A. \& Cattuto, C.
in \textit{Social Phenomena,
(eds  Gon\c{c}alves, B. and Perra, N.)
Ch 3, 37-57 (Springer International Publishing Switzerland, 2015).}

\bibitem{Toth:2015}
Toth, D. J. A.  et al.
The role of heterogeneity in contact timing and duration in network models of influenza spread in schools.
{\em J. Roy. Soc. Int.} {\bf 12,} 20150279 (2015).

\bibitem{Guclu:2016}
Guclu H. et al. 
Social Contact Networks and Mixing among Students in K-12 Schools in Pittsburgh, PA.
{\em PLoS ONE} {\bf 11,} e0151139 (2016).

\bibitem{Smieszek:2014} 
Smieszek, T. et al.
How should social mixing be measured: comparing web-based survey and sensor-based methods. 
{\em BMC Infect. Dis.} {\bf 14,} 136 (2014). 

\bibitem{Smieszek:2016} 
Smieszek, T. et al.
Contact diaries versus wearable proximity sensors in measuring contact patterns at a conference: method comparison and participants' attitudes
{\em BMC Infect. Dis.} {\bf 16,} 341 (2016).


\bibitem{Stopczynski:2015}
Stopczynski, A., Sapiezynski, P. \&  Lehmann, S. 
Temporal fidelity in dynamic social networks.
{\em Eur. Phys. J.} {\bf 88,} 249 (2015). 

\bibitem{Lee:2006}
Lee, S. H., Kim, P.-J., Jeong, H. Statistical properties of sampled networks.
{\em Phys. Rev. E} {\bf 73,} 016102 (2006).

\bibitem{Kossinets:2006}
\bibinfo{author}{Kossinets, G.}
\newblock \bibinfo{title}{Effects of missing data in social networks}.
\newblock \emph{\bibinfo{journal}{Social Networks}}
  \textbf{\bibinfo{volume}{28}}, \bibinfo{pages}{247 -- 268}
  (\bibinfo{year}{2006}).

\bibitem{Ghani:1998}
\bibinfo{author}{Ghani, A.~C.}, \bibinfo{author}{Donnelly, C.~A.} \&
  \bibinfo{author}{Garnett, G.~P.}
\newblock \bibinfo{title}{Sampling biases and missing data in explorations of
  sexual partner networks for the spread of sexually transmitted diseases}.
\newblock \emph{\bibinfo{journal}{Statistics in Medicine}}
  \textbf{\bibinfo{volume}{17}}, \bibinfo{pages}{2079--2097}
  (\bibinfo{year}{1998}).

\bibitem{Genois:2015} 
G\'enois, M., Vestergaard, C., Cattuto, C. \& Barrat, A.
Compensating for population sampling in simulations of epidemic spread on temporal contact networks.
{\em Nat. Comm.} {\bf 6,} 9860 (2015).


\bibitem{Vestergaard:2016}
Vestergaard, C., Valdano, E., G\'enois, M., Poletto, C., Colizza, V. \& Barrat, A.
Impact of spatially constrained sampling of temporal contact networks on the evaluation of the epidemic risk 
{\em European Journal of Applied Mathematics} {\bf 27,} 941 (2016).

\bibitem{Bliss:2014}
\bibinfo{author}{Bliss, C.~A.}, \bibinfo{author}{Danforth, C.~M.} \&
  \bibinfo{author}{Dodds, P.~S.}
\newblock \bibinfo{title}{Estimation of global network statistics from
  incomplete data}.
\newblock \emph{\bibinfo{journal}{PLoS ONE}} \textbf{\bibinfo{volume}{9}},
  \bibinfo{pages}{e108471} (\bibinfo{year}{2014}).

\bibitem{Zhang:2015}
Zhang, Y., Kolaczyk, E. D. \& Spencer, B. D. 
Estimating network degree distributions under sampling: An inverse
problem, with applications to monitoring social media
networks. {\em Ann. Appl. Stat.} {\bf 9,} 166 (2015).

\bibitem{Squartini:2017}
Squartini, T., Cimini, G., Gabrielli, A. \& Garlaschelli, D.
 Network reconstruction via density sampling.
{\em App. Netw. Sci.} {\bf 2,} 3 (2017).

\bibitem{Mastrandrea:2016}
Mastrandrea, R. \& Barrat, A.
How to estimate epidemic risk from incomplete contact diaries data?
{\em PLoS Comput Biol} {\bf 12,} e1005002 (2016).

\bibitem{Fournet:2016}
Fournet, J. \& Barrat, A.
Epidemic risk from friendship network data: an equivalence with a non-uniform sampling of contact networks.
{\em Sci. Rep.} {\bf 6,} 24593 (2016).

\bibitem{Stehle:2011} 
Stehl\'e, J. et al.
Simulation of an SEIR infectious disease model on the dynamic contact network of conference attendees. 
{\em BMC Med.} {\bf 9,} 87 (2011). 

\bibitem{Machens:2013}
Machens, A. et al.
An infectious disease model on empirical networks of human contact: bridging the gap between dynamic network data and contact matrices 
{\em BMC Infectious Diseases} {\bf 13,} 185 (2013).

\end{thebibliography}
\end{document}